\documentclass[sn-mathphys-num]{sn-jnl}

\usepackage{graphicx}%
\usepackage{multirow}%
\usepackage{amsmath,amssymb,amsfonts,bbm}%
\usepackage{amsthm}%
\usepackage{mathrsfs}%
\usepackage[title]{appendix}%
\usepackage{xcolor}%
\usepackage{textcomp}%
\usepackage{manyfoot}%
\usepackage{booktabs}%
\usepackage{algorithm}%
\usepackage{algorithmicx}%
\usepackage{algpseudocode}%
\usepackage{listings}%

\begin{document}

\title[Article Title]{Effective Affinity for Generic Currents in Markov Processes}

\author*[1]{\fnm{Adarsh} \sur{Raghu}}\email{adarsh.raghu@kcl.ac.uk}

\author[1]{\fnm{Izaak} \sur{Neri}}\email{izaak.neri@kcl.ac.uk}

\affil[1]{\orgdiv{Department of Mathematics}, \orgname{King's College London}, \orgaddress{\street{The Strand}, \city{London}, \postcode{WC2R2LS},  \country{United Kingdom}}}

\abstract{In nonequilibrium systems with uncoupled currents, the thermodynamic affinity determines the direction of  currents, quantifies dissipation, and constrains current fluctuations. However, these properties of the thermodynamic affinity do not hold in complex systems with multiple coupled currents.   For this reason, there has been an ongoing search in nonequilibrium thermodynamics for an affinity-like quantity, known as the effective affinity, which applies to a single current in a system with multiple coupled currents.    Here, we introduce an effective affinity that applies to  generic currents in time-homogeneous Markov processes. We show that the effective affinity is a single number  encapsulating     several dissipative and fluctuation properties of fluctuating currents: the effective affinity determines the direction of flow of the current;   the effective affinity  multiplied by the current   is a lower bound for  the rate of dissipation; for  systems with uncoupled currents the effective affinity equals the standard thermodynamic affinity; and  the  effective affinity constrains negative fluctuations of currents, namely, it is the exponential decay constant of the distribution of current infima.     We derive the above properties with large deviation theory and martingale theory, and one particular interesting finding is a class of martingales associated with generic currents.  Furthermore, we make a study of the relation between effective affinities and stalling forces in a biomechanical model of motor proteins, and we find that both quantities are approximately equal  when this particular model is thermodynamically consistent.     This brings interesting perspectives on the use of stalling forces for the estimation of dissipation. }


\maketitle

\section{Introduction}
\label{sec:intro}

The term affinity originates  in chemistry, where  it was used to describe the force that drives a chemical reaction  (originally as vague as in the sense of the attraction between particles~\cite{as2010affinity}).       Its precise thermodynamic formulation as a  chemical potential difference  originates in the works of  Th\'{e}ophile de Donder~\cite{affinite, de1936thermodynamic,rocci2024celebrating} on irreversible chemical reactions.   In particular,  De Donder demonstrates the inequality
\begin{equation}
\dot{s} = \sum_{\gamma\in \mathcal{F}} \frac{\mathsf{a}_\gamma}{k_{\rm_B}\mathsf{T}_{\rm env}} \overline{j}_\gamma \geq 0  ,\label{eq:sdotMacro}
\end{equation}
where $\gamma$ is an index labeling the  different chemical reactions that can take place, $\mathcal{F}$ represents the set of all these reactions,  $\overline{j}_{\gamma}$ is the  velocity of the $\gamma$-th reaction,  $\mathsf{a}_{\gamma}$ is the corresponding affinity, $k_{\rm_B}$ is the Boltzmann constant, and  $\mathsf{T}_{\rm env}$ is the temperature;  see   Chapter 4 in Ref.~\cite{bookPrigo} for a more extensive discussion.    
Equation (\ref{eq:sdotMacro}) expresses   the rate of entropy production  $\dot{s}$ (divided by $k_{\rm B}$) as a sum over the  forces $\mathsf{a}_\gamma/(k_{\rm_B}\mathsf{T}_{\rm env})$ multiplied by their corresponding currents $\overline{j}_\gamma$.  Such decompositions of  entropy production  in terms of forces and fluxes apply to a large number of nonequilibrium systems~\cite{bookPrigo}.   

If the reactions in (\ref{eq:sdotMacro})  are uncoupled, then the affinities $a_{\gamma}$ specify the sign of the velocities $\overline{j}_\gamma$.   Thus,  affinities can be seen as the driving forces that determine the directions of  (uncoupled) irreversible chemical reactions.  However, in complex systems consisting of a large number of  coupled chemical reactions, as occurs for instance in metabolic pathways of living cells,  the inequality (\ref{eq:sdotMacro}) is a weak constraint that does not determine the  direction of the individual  reactions~\cite{bookPrigo}.   Coupling can also occur between chemical and mechanical fluxes, as is  the case for self-propelled motion  molecular motors~\cite{astumian1996mechanochemical, Liepelt}.  
Building on recent advances in nonequilibrium thermodynamics, that we briefly summarise next,  we develop here an effective affinity that serves as generalisation of affinity in    systems with coupled currents.

In the 1970s, a mesoscopic theory for entropy production  in nonequilibrium systems, including irreversible chemical reactions, was developed based on the theory of Markov processes (see Refs.~\cite{hill2013free, schnakenberg}, or Chapter 3 of~Ref.\cite{peliti2021stochastic} for a recent review).   Consider a Markov chain with transition  rates that satisfy the local detailed balance condition~\cite{maes2021local,peliti2021stochastic}.  The entropy production (\ref{eq:sdotMacro}) is then described by the Schnakenberg formula~\cite{schnakenberg, peliti2021stochastic} 
\begin{equation}
\dot{s}  = \sum_{\gamma\in\mathcal{C}}a_\gamma \overline{j}_\gamma, \label{eq:meso}
\end{equation} 
where now $\gamma$ is an index that labels a basis of fundamental cycles in the graph of admissible transitions of the underlying Markov chain, where $\overline{j}_\gamma$  is the  current associated with the cycle $\gamma$, and $a_\gamma $ is the  corresponding cycle affinity.  Importantly,  all quantities in Eq.~(\ref{eq:meso})  admit an explicit expression in terms of the transition rates of the Markov chain~\cite{schnakenberg}, and hence $\dot{s}$ is a function of the transition rates of the Markov chain.  The  number of terms in Eq.~(\ref{eq:meso}) is in general significantly larger than the number of terms in Eq.~(\ref{eq:sdotMacro}), as the macroscopic currents in (\ref{eq:sdotMacro}) can be expressed as a linear combination of  a multiple of the cycle currents in Eq.~(\ref{eq:meso}).      Notice that for isothermal systems,  $\mathsf{a}_\gamma = a_\gamma/(k_{\rm_B}\mathsf{T}_{\rm env})$, relating the ``de Donder" affinity $\mathsf{a}_{\gamma}$ in Eq.~(\ref{eq:sdotMacro}) to the Markov chain affinity  $a_{\gamma}$  of Eq.~(\ref{eq:meso}). However,   the affinities  $a_{\gamma}$ also apply to systems in contact with multiple thermal reservoirs~\cite{gaspard2013multivariate}.  

Currents in Markov processes have fluctuations, and the affinities $a_\gamma$   play a role in quantifying the nonequilibrium fluctuations of currents.   As was discovered in the 1990s~\cite{lebowitz1999gallavotti}, fluctuating integrated currents $J_\gamma$ associated with the average currents $\overline{j}_\gamma$    satisfy   fluctuation relations~\cite{seifert2005, andrieux2007fluctuation},  including the integral relation
\begin{equation}
\Big\langle e^{- \sum_{\gamma \in \mathcal{C}}a_{\gamma}J_\gamma}
\Big\rangle = 1 ,\label{eq:IFT}
\end{equation}
where $\langle \cdot\rangle$ denotes an average over repeated realisations of the process. 
Applying Jensen's inequality to (\ref{eq:IFT}) and identifying $\langle J_\gamma\rangle = \overline{j}_\gamma t$ yields the second law of thermodynamics  $\dot{s}\geq 0$, and fluctuation relations can thus be viewed as generalizations of the second law of thermodynamics

For  nonequilibrium processes  that have one macroscopic current, the affinity $a$ (we dropped here the $\gamma$ index as there is only one current) captures  both dissipative and fluctuation properties of the fluctuating current.  Indeed,  the second law of thermodynamics, $\overline{s}=\overline{j}a>0$, implies that the sign of the affinity $a$ equals the sign of the average current  $\overline{j}$.    In addition, the integral fluctuation relation $\langle e^{-aJ}\rangle = 1$ implies that fluctuations of $J$ that have opposite sign of $\overline{j}$ are exponentially constrained, i.e.,  the probability of observing a value $J_t<j$ is smaller or equal than $\exp(-|aj|)$~\cite{seifert2012stochastic, jarzynski2012equalities}.  

For systems with multiple coupled currents, the second law of thermodynamics does not determine the sign of the current from the affinity, and for multiple currents it is also unclear how   the fluctuation relation (\ref{eq:IFT})   constrains  the fluctuations of an individual current.     Hence,  although  thermodynamic affinities  determine the constraints imposed by the second law of thermodynamics on a system's dynamics, this constraint is   in general not   useful for quantifying properties of a single fluctuating current of interest;   for example, thermodynamic affinities do not determine the sign of a current in systems with multiple coupled reactions.  
This brings us to the concept of an  {\it effective affinity}, which is an  affinity-like number  that encapsulates a large number of interesting properties of fluctuating currents, even in systems with multiple currents.       

The  effective affinity problem is the following.   
Assume we observe a single fluctuating current $J_t$ in a nonequilibrium system that has  multiple  currents, i.e., $|\mathcal{C}|>1$.  Is there a natural way of assigning an {\it effective} affinity $a^\ast$ to the observed current that captures the nonequilibrium properties of the current as seen from an observer that  only measures this current?    The effective affinity should have properties that physicists and chemists    associate to affinities of currents in systems with one current.  For example, the effective affinity should determine the direction of flow of the current, quantify  the rate of dissipation, and it should constrain the  fluctuations of the current against its average flow. In systems with one current  the thermodynamic affinity has all the desired properties, and thus the effective affinity should equal the thermodynamic affinity in those cases.      On the other hand,  for systems with multiple  coupled currents, the  effective affinity  is  in general different from the  current's corresponding thermodynamic affinity.    

The effective affinity problem has appeared a few times before in the literature, and we provide here, to the best of our knowledge,  an overview of previous work.    The effective affinity was first studied in terms of the fluctuation properties of currents.  
Building on Ref.~\cite{andrieux2007network}, the paper ~\cite{gaspard2013multivariate} defines an effective affinity for    strongly coupled currents.   Strongly coupled currents  satisfy asymptotically in the limit of large times a detailed fluctuation relation.  In this case, the effective affinity is  the prefactor that appears in front of the current in the exponential function that appears in the fluctuation relation.     Analogously, in Ref.~\cite{bulnes2013effective}   the effective affinity is defined for an electron transport problem through the integral fluctuation relation.     Note that these references define the effective affinity through fluctuation relations, but additional properties of this quantity are not explored.    This changed with the recent works~\cite{polletini1, bisker2017hierarchical,polettini2019effective, neri2023extreme} that develop a theory for effective affinities  of  {\it edge currents}.   
These are currents that monitor the net number of transitions along a single edge of the graph of admissible transitions in  a Markov chain.   For edge currents, the effective affinity satisfies asymptotically an integral fluctuation relation and therefore it specifies the direction of the current, as in Refs.~\cite{gaspard2013multivariate, bulnes2013effective}.  However, two new important properties were found.  First, it was found that  the effective affinity multiplied by the current is a lower bound for the rate of dissipation~\cite{bisker2017hierarchical, polettini2019effective}, and second it was shown that the effective affinity is the additional force required to stall the current~\cite{polletini1}.   Further works show that the effective affinity for edge currents defines a martingale process~\cite{neri2023extreme},  determines the extreme value statistics of currents~\cite{neri2023extreme}, and that for unicyclic systems the edge effective affinity equals the thermodynamic affinity~\cite{neri2023extreme}.    Hence,  in conclusion, for the specific class of edge currents   these papers demonstrate  that the effective affinity encapsulates a large number of  properties of fluctuating currents, and these are properties   that physicists associate with affinities of currents in unicyclic systems.   

So far, the question remained open whether effective  affinities can be defined for {\it generic} fluctuating currents in Markov processes.     This question is a crucial step forward in resolving the  applicability of the effective affinities, as fluctuating currents in experiments, such as   position of a molecular motor or the beat of a cilia~\cite{battle2016broken, di2024variance}, cannot be assumed to be edge currents.     

In this Paper, we  define an effective affinity  for {\it generic} currents in a Markov process, addressing the issues with the general applicability of the effective affinity.   We demonstrate that the effective affinity   inherits all the properties of the edge effective affinity, except  for the stalling force property; nevertheless, we  derive  results for the stalling force in  a case study of a molecular motor that raise interesting perspectives on the relation between effective affinity and stalling forces for future research.  

We summarise with more detail the properties of the effective affinity.   
Given a   fluctuating current $J_t$ in a Markov process $X_t$, we define the  effective affinity $a^\ast$ for generic currents through the asymptotic integral fluctuation relation

\begin{equation}
\lim_{t\rightarrow \infty} \frac{1}{t} \ln \Big\langle e^{-a^\ast J_t}\Big\rangle = 0 ,\label{eq:integral}
\end{equation} 

where $\langle \cdot\rangle$ is an average over repeated realisations of the process;  note that for currents with nonzero average values, $\langle J_t \rangle \neq 0$,  
  Eq.~(\ref{eq:integral})  has at most one unique nonzero  solution.  If   the current is the stochastic entropy production~\cite{Seki, seifert2012stochastic}, then according to the Gallavotti-Cohen fluctuation relation~\cite{lebowitz1999gallavotti}, $a^\ast=1$, and  for edge currents that count the number of transitions along a single edge of the graph of admissible transitions of a Markov jump process, $a^\ast$  equals the effective  affinity as developed in Refs.~\cite{polletini1,bisker2017hierarchical,polettini2019effective}. 

We outline the main properties that characterize $a^\ast$ as an effective affinity. 
Applying Jensen's inequality  to (\ref{eq:integral}), we  obtain
 \begin{equation}
 a^\ast \overline{j}\geq 0, \label{eq:sign}
 \end{equation}
 and thus the effective affinity determines the sign of the corresponding current.   Hence, from (\ref{eq:sign}) it follows that the effective affinity can be seen as a force that drives a current.

As a second key property, we find that the  effective affinity quantifies the dissipation contained in a  current.   Specifically, using large deviation theory we find that 
\begin{equation}
a^\ast \overline{j} \leq \dot{s}, \label{eq:lowerbound}
\end{equation} 
where $\overline{j} = \lim_{t\rightarrow \infty}\langle J_t\rangle/t$ is the average current associated with the observed current $J_t$, and  
where $\dot{s}$ is the  average rate of dissipation. The inequality (\ref{eq:lowerbound})  is suggestive of the equalities  (\ref{eq:sdotMacro})  and (\ref{eq:meso}).  However, since the effective affinity applies to a single current, it   captures in general a portion of the total dissipation, as expressed by the inequality (\ref{eq:lowerbound}).

Thirdly, we show that the effective affinity   constrains fluctuations of generic currents.  Let us assume that $\overline{j}>0$ so that we can define the infimum  value 
\begin{equation}
J_{\rm inf} := {\rm inf}\left\{J_t:t\geq 0\right\}.
\end{equation}
  It then holds that the tails of the distribution $p_{J_{\rm inf}}$ of  $J_{\rm inf}$  are exponential with a  decay constant  given by  $a^\ast$, i.e., 
\begin{equation}
p_{J_{\rm inf}}(j)  = \exp\left(a^\ast j[1+o_j(1)]\right),\quad j\leq 0,  
\label{eq:inf}
\end{equation}
where $o_j(1)$ is a function that vanishes in the limit of large $|j|$.  
The extreme value law (\ref{eq:inf})  extends the exponential law  for the infimum statistics  of entropy production, see  Refs.~\cite{neri2017statistics, neri2019integral},  to  generic currents in Markov processes .

As a  fourth key property, we find that  the effective affinity $a^\ast$ equals the thermodynamic affinity $a_{\gamma}$ when the $|\mathcal{C}|$ currents in Eq.~(\ref{eq:meso}) are uncoupled.   This result confirms that the effective affinity is a proper extension of the thermodynamic affinity to cases where the currents are coupled.

Apart from these key properties, we also investigate a few other features of the effective affinity. Notably, we identify a set of optimal currents that attain the equality in (\ref{eq:lowerbound}), and we investigate the relationship between effective affinities and  the current's  stalling force.  The stalling force is the the additional force required to stall a current.   Since the stalling force equals the effective affinity if a current  captures the transitions along a single edge of a Markov chain (i.e., it is an edge current,  see Ref.~\cite{polettini2019effective}),  we suspect that there may be a similar connection between stalling forces and effective affinities for generic currents.   In  this Paper we make a numerical study of effective affinities and  stalling forces  in a biophysical model of a molecular motor, and we find that they are not equal to each other for general currents.  Nevertheless, when this particular model is thermodynamically consistent, then the stalling force is almost equal to the effective affinity (with a relative error of the order $10^{-2}$ in the studied example), and hence in this example the   stalling force can be used as an estimate for the effective affinity.

From a mathematical point of view, the effective affinity relates  various concepts, including, large deviation theory, martingale theory, and extreme value statistics and splitting probabilities of currents.  In large deviation theory, the effective affinity is the nonzero root of the  logarithmic moment generating function, and  in martingale theory  the effective affinity appears as the prefactor in front of the current in the expression of an exponential martingale.   Note that these exponential martingales generalise martingales studied previously in literature, such as,  the exponentiated negative fluctuating entropy production \cite{chetrite2011two, neri2017statistics} and an exponential martingale related to edge currents~\cite{neri2023extreme}.     In this Paper, we use  exponential  martingales of generic currents to determine the   splitting probabilities  of currents in first-passage problems with two boundaries.   Furthermore, we use   exponential martingales to derive a thermodynamic inequality involving first-passage quantities that expresses a trade off between speed, dissipation, and uncertainty~\cite{neri2022universal}.

The paper is structured as follows:
In Sec.~\ref{sec:syssetup}, we describe the general setup of Markov jump processes that we will be working with for most of the paper. In Sec.~\ref{sec:PresentingEA}, we define the effective affinity with large deviation theory, and we  use large deviation theory to derive the   bound~(\ref{eq:lowerbound}). In Sec.~\ref{sec:martingale}, we relate the effective affinity with martingale theory by defining a martingale  associated with a  generic fluctuating  current, and in Sec.~\ref{sec:fpp} we use martingale theory to  derive the infimum law \eqref{eq:inf}, linking the effective affinity with the current's extreme value statistics.   In Sec.~\ref{sec:tightness}, we derive a sufficient condition for optimal curents (these are currents that attain the equality in (Eq.~\ref{eq:lowerbound})) and  this leads us to the concept of cycle equivalence classes.  In this section we also show the equivlence between effective affinities and thermodynamic affinities of uncoupled currents.   In the following three sections, we will explore questions related to effective affinity that are not generically  addressed in this paper. Instead, we will examine these problems through  case studies.
In  Sec.~\ref{sec:twocycle}, we numerically investigate the optimal currents in a  simple model with two cycles with the aim of identifying   necessary conditions for  current optimality.      In Sec.~\ref{sec:diffusion}, we make a brief detour in Markov processes on continuous state spaces.  We explore the effective affinity for a particle subject to a constant force and  undergoing overdamped diffusion.    In Sec.~\ref{sec:kinesin} we analyse the effective affinity in a  biochemical model of the molecular motor Kinesin-1,  and we demonstrate that the   stalling force approximates well the effective affinity when the model is thermodynamically consistent.   We end the paper with a Discussion in Sec.~\ref{sec:discussion} and four appendices with technical details.

\section{Fluctuating  currents in  Markov jump processes and entropy production}
\label{sec:syssetup}
For simplicity, we focus in this Paper  on Markov jump processes in a discrete state space.
We consider time-homogeneous Markov jump processes $X_t\in \mathcal{X}$ on a finite set   $\mathcal{X}$, which are  defined by their $\mathbf{q}$-matrix~\cite{norris1998markov, liggett2010continuous}.  The  off-diagonal entries $\mathbf{q}_{xy}$  denote the rate at which $X_t$ jumps from $x$ to $y$, with $x,y \in \mathcal{X}$.          The diagonal entries $\mathbf{q}_{xx} = -\sum_{y\in \mathcal{X}\setminus \left\{x\right\}}\mathbf{q}_{xy}$ denote the exit rates out of the state $x$.        The probability mass function $p_t(x)$ of $X_t$ solves the differential equation 
\begin{equation}
\partial_t p_t(x) = \sum_{y\in \mathcal{X}}p_t(y)\mathbf{q}_{yx} . \label{eq:9}
\end{equation} 
The stationary state $p_{\rm ss}(x)$ is the left eigenvector of $\mathbf{q}$ associated with its Perron root (the Perron root is the eigenvalue with the largest real part).  We assume that $X_t$ is ergodic, so that $p_{\rm ss}$ is unique and $p_{\rm ss}(x)>0$~\cite{bremaud2013markov}, and we also assume that transitions are reversible, i.e., $\mathbf{q}_{xy}>0$ then also $\mathbf{q}_{yx}>0$.  

Fluctuating integrated currents $J_t$ are time-additive and time-reversal antisymmetric, fluctuating observables. Any such observable defined on a Markov jump process $X_t$ can be expressed as a linear combination
\begin{equation}
J_t := \frac{1}{2}\sum_{x,y\in \mathcal{X}}c_{x y}J^{x y}_t \label{eq:genericCurrent}
\end{equation}
of edge currents $J^{x y}_t$, which are defined as the difference between the number of forward jumps $N^{x y}_t $ and the number of backward jumps $N^{y x}_t$ between $x$ and $y$ in the interval $[0,t]$, i.e.,
\begin{equation}
J^{x y}_t := N^{x y}_t - N^{y x}_t.\label{eq:jxyt}
\end{equation}
The coefficients $c_{x y}=-c_{y x}\in\mathbb{R}$  quantify the amount of resource transported   when the process jumps from $x$ to $y$.  For example, the coefficients $c_{xy} $ may  denote the amount of energy exchanged with a thermal reservoir, the number of particles exchanged with external chemostats, or the positional distance covered when the system moves from $x$ to $y$.       If $\mathbf{q}_{xy} = 0$, then the corresponding    coefficient $c_{xy}$   is irrelevant.   Therefore, the relevant coefficients $c_{xy}$  span an Euclidean space of dimension $|\mathcal{E}|$, where $\mathcal{E}$ is the set of nondirected edges in the graph of admissible transitions (i.e., the pairs $(x,y)$  with $\mathbf{q}_{xy}\neq 0$). The corresponding average current  $\overline{j}$ takes the expression
\begin{equation}
\overline{j} := \lim_{t\rightarrow \infty} \langle J_t\rangle/t = \frac{1}{2}\sum_{x\in \mathcal{X}}\sum_{y\in \mathcal{X}\setminus \left\{x\right\}}c_{xy} \overline{j}_{xy}, \label{eq:javgsum} 
\end{equation}
where  
\begin{equation}
\overline{j}_{xy} := \lim_{t\rightarrow \infty} \langle J^{xy}_t\rangle/t = p_{\rm ss}(x)\mathbf{q}_{xy}-p_{\rm ss}(y)\mathbf{q}_{yx} \label{eq:currentjxy}
\end{equation}
is the average of the edge current  associated with the transition between states $x$ and $y$.  Without loss of generality, we assume in this Paper that $\overline{j}>0$.  

An important example of a fluctuating current is the fluctuating entropy production~\cite{Seki, seifert2012stochastic, peliti2021stochastic},
\begin{equation}
S_t := \frac{1}{2} \sum_{x\in \mathcal{X}}\sum_{y\in \mathcal{X}\setminus \left\{x\right\}} J^{x y}_t \ln \frac{p_{\rm ss}(x)\mathbf{q}_{xy}}{p_{\rm ss}(y)\mathbf{q}_{yx}}, \label{eq:St}
\end{equation} 
which is the fluctuating current with coefficients $c_{xy}$  that are equal to the microscopic edge affinities $c_{xy} =\ln \frac{p_{\rm ss}(x)\mathbf{q}_{xy}}{p_{\rm ss}(y)\mathbf{q}_{yx}}$.  Using the principle of local detailed balance, we can identify the average rate
\begin{equation}
\dot{s} := \lim_{t\rightarrow \infty}\langle S_t\rangle/t = \frac{1}{2} \sum_{x\in \mathcal{X}}\sum_{y\in \mathcal{X}\setminus \left\{x\right\}} \overline{j}_{xy} \ln \frac{p_{\rm ss}(x)\mathbf{q}_{xy}}{p_{\rm ss}(y)\mathbf{q}_{yx}}
\end{equation}
with the rate of dissipation~\cite{maes2021local}. 
The rate of dissipation $\dot{s}$ can also be expressed as a sum of the form  Eq.~(\ref{eq:meso}), where  $\gamma$ label the cycles in a fundamental cycle basis $\mathcal{C}$  of the graph of admissible transitions, where  $\overline{j}_{\gamma}$ is the current along an edge of the cycle $\gamma$, and where 
\begin{equation}
a_{\gamma} = \ln \frac{\prod_{(x,y)\in \gamma}\mathbf{q}_{xy}}{\prod_{(x,y)\in -\gamma}\mathbf{q}_{xy}}  \label{eq:agamma}
\end{equation}
are the cycle affinities;  we have labeled with $-\gamma$   the cycle obtained from $\gamma$ by changing the orientation of the edges (see Sec.~\ref{sec:recap} for details).

\section{Effective affinity from large deviations of currents}\label{sec:PresentingEA}
We initiate our study of the effective affinity in large deviation theory, where it naturally appears as the non-zero root of the logarithmic moment generating function.      

\subsection{Definition}
Currents of the form \eqref{eq:genericCurrent} satisfy a large deviation principle when $\overline{j}\neq 0$~\cite{maes2008steady,maes2008canonical,bertini2015flows}.   Indeed, the probability distribution of $J_t/t$ takes for large values of $t$ the form 
\begin{equation}
    p_{J_t/t}\left(j\right) = \exp\left(-t \mathcal{I}(j)[1+o_t(1)]\right),
\end{equation}
where $\mathcal{I}(j)$ is the rate function of $J_t$ and $o_t(1)$ denotes an arbitrary function that converges to zero for large values of $t$. According to the G\"{a}rtner-Ellis theorem~\cite{touchette2009large, dembo2009large}, $\mathcal{I}(j)$ is  the Legendre-Fenchel transform   of the logarithmic moment generating function 
\begin{equation}
\lambda_J(a) := \lim_{t\rightarrow \infty}\frac{1}{t} \ln \Big\langle  e^{-aJ_t}\Big\rangle  \label{eq:deflambda}
\end{equation} 
such that  
\begin{equation}
    \mathcal{I}_J(j) ={\rm max}_{a\in \mathbb{R}}(-\lambda_J(a)-aj).
\end{equation}

\begin{figure}[t!]\centering
 \includegraphics{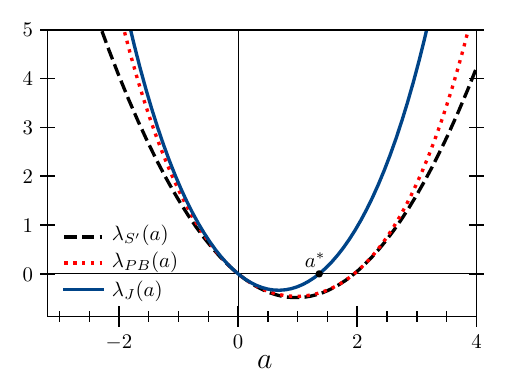}
\caption{{\it Illustrated definition of the effective  affinity $a^\ast$}.   The logarithmic moment generating function $\lambda_J$  is plotted as a function of $a$   for an example current $J$ in the four state model of Fig.~\ref{fig:3} (see Appendix~\ref{sec:fig1params} for  definitions).  The effective affinity $a^\ast$, defined as the non-zero root of $\lambda_J(a)$, is marked by a circle.      In addition,  the logarithmic moment generating function $\lambda_{S'}$ for  the rescaled fluctuating entropy production $S'=S/\dot{s}$ , and the parabola $\lambda_{PB}(a) = a \overline{j}(-1+a \overline{j}/\dot{s})$  that appears on the right hand side of the inequality (\ref{eq:bound2}) are plotted as a function of $a$.  Note that  in both cases ($J_t$ and $S'_t$) the average current equals $1$.      }\label{fig:1}   
\end{figure}

We define the effective affinity $a^\ast$ as the nonzero root of $\lambda_J(a)$ (see  Fig.~\ref{fig:1} for an illustration), i.e.,
\begin{equation}
\lambda_J(a^\ast) = 0, \label{eq:defa}
\end{equation} 
a definition that is consistent with Eq.~(\ref{eq:integral}).   If $\overline{j} = 0$, then $\lambda_J(a)$ has no nonzero root, and  we set $a^\ast=0$.   
Notice that  the cumulants of $J_t/t$ are determined by the derivatives of $\lambda_J(a)$ at the zero root, $a=0$. 
  Instead in this paper, we highlight the importance of the  second root $a^\ast$ of the logarithmic moment generating function that captures atypical properties of $J$.

  Applying Jensen's inequality to Eq.~(\ref{eq:integral}) we find 
\begin{equation}
a^\ast \overline{j} \geq 0.  
\end{equation} 
Hence,  $a^\ast$ has the same sign as $\overline{j}$ and therefore we can say that the effective affinity sets the direction in which the current flow.  Note that  this property does not hold for  the thermodynamic affinities in Eq.~(\ref{eq:sdotMacro}), and  the effective affinity is thus in general different from the  thermodynamic affinity.   

\subsection{Lower bound on dissipation}
The effective affinity multiplied by the average current captures a portion of the entropy production rate of the system as expressed by the inequality  Eq.~(\ref{eq:lowerbound}), i.e., $\dot{s}\geq a^\ast \overline{j}$.     This is one of the main properties of the effective affinity, and we derive it here from large deviation theory. 

 The inequality  (\ref{eq:lowerbound})  follows from the universal lower bound 
 \begin{equation}
 \lambda_J(a) \geq a \overline{j}\left(-1+a\frac{\overline{j}}{\dot{s}}\right),\label{eq:bound2}
 \end{equation}  
on the logarithmic moment generating function of $J_t$.      The parabolic bound (\ref{eq:bound2})  was conjectured in \cite{pietzonka}  for arbitrary currents   $J_t$ in Markov jump processes based on substantial numerical evidence, and subsequently derived in Ref.~\cite{gingrich2016dissipation}  by applying  the contraction principle to  the level 2.5 large deviation rate function.    Since the effective affinity $a^\ast$ is the positive  root of $ \lambda_J(a)$, it is smaller or equal than the  positive root  $a=\dot{s}/\overline{j}$  of the  right-hand side of (\ref{eq:bound2}), as illustrated in Fig.~\ref{fig:1}, which  concludes the derivation of the bound \eqref{eq:lowerbound}. 

\subsection{Limiting cases}\label{subsec:EALimiting}
We consider the effective affinity in two relevant limiting cases of the current $J_t$.  If $J_t = J_t^{xy}$, then 
the effective affinity equals (see Appendix~\ref{sec:S6}),  
\begin{equation}
a^\ast = \ln \frac{p^{(x,y)}_{\rm ss}(x)\mathbf{q}_{xy}}{p^{(x,y)}_{\rm ss}(y) \mathbf{q}_{yx}} \label{eq:effectEdge}
\end{equation}
where $p^{(x,y)}_{\rm ss}(x)$ is the probability mass function of a modified Markov jump process for which the transition rates along the  $(x,y)$-edge have been set to zero.   On the right-hand side of Eq.~(\ref{eq:effectEdge}) we recognise the the effective affinity for edge currents as studied in Refs.~\cite{polletini1,bisker2017hierarchical,polettini2019effective}, and hence  $a^\ast$ extends the effective edge affinity (\ref{eq:effectEdge})  to generic currents in Markov processes.

Another interesting limiting case is when $J_t = kS_t$ with $k\in \mathbb{R}$.        Such currents satisfy the Gallavotti-Cohen  symmetry~\cite{lebowitz1999gallavotti} 
 \begin{equation}
 \lambda_{J}(a)  =  \lambda_J\left(k^{-1} -a\right), \label{eq:galavottiCohen}
 \end{equation}
and thus  $a^\ast = 1/k$. 
In addition, since $\overline{j} = k\dot{s}$,  the equality in (\ref{eq:lowerbound}) is  attained~\cite{neri2022universal} for  currents that are proportional to the stochastic entropy production.   

There is another interesting limiting case corresponding with Markov processes that contain unidirectional transitions, such that there exist a pair $(x,y)$ for which  $\mathbf{q}_{xy}>0$ but $\mathbf{q}_{yx}=0$; notice that we have excluded such cases in  the original setup.  If the graph of admissible transitions is unicyclic, then the scaled cumulant generating function $\lambda_J(a)$ does not have a nonzero root.   This situation corresponds with an infinitely large effective affinity, as $a^\ast$ diverges when $\mathbf{q}_{yx}\rightarrow 0$.  Correspondingly, the rate of dissipation $\dot{s}$ diverges in this limit.  

\subsection{Effective affinity from the tilted generator}
For Markov processes defined on finite sets, 
 we can readily obtain $\lambda_J(a)$, and thus also the effective affinity $a^\ast$, from the spectrum of a ``tilted" $\mathbf{q}$-matrix. 
Indeed,  applying Kolmogorov's backward  equation to $\langle e^{-aJ_t}\rangle$,   it follows that  $\lambda_J(a)$ 
is the Perron root of the tilted matrix~\cite{lebowitz1999gallavotti, touchette2009large,  chetrite_nonequilibrium_2015, baiesi2009computation, jack2010large}
\begin{equation}
\tilde{\mathbf{q}}_{xy}(a) :=  \left\{\begin{array}{ccc}  \mathbf{q}_{xy}e^{-a c_{x y}}, &{\rm if}& x\neq y, \\  -\sum_{z\in \mathcal{X}\setminus \left\{x\right\}} \mathbf{q}_{xz},&{\rm if}&   x=y.  \end{array}\right. \label{eq:qtilde}
\end{equation}   
The effective affinity $a^\ast$ is thus the value of $a$ at which the Perron root vanishes, and obtaining $a^\ast$ is thus a straightforward computation  when the set  $\mathcal{X}$ is finite.   

\section{Effective affinity from martingale theory} \label{sec:martingale}
Let $M_t$ be a stochastic process that is a function of the trajectory of $X$ in the interval $[0,t]$. The process $M_t$ is a martingale if it has 
no drift, i.e.,
\begin{equation}
\langle M_t|X^s_0 \rangle = M_s \label{eq:mart}
\end{equation}
for all $s\in [0,t]$, where $\langle \cdot|X^s_0\rangle$ denotes the expectation conditioned on the  trajectory $X^{s}_0 := \left\{X_u: u\in [0,s]\right\}$.    

We show in this section that the process 
\begin{equation}
M_t := \phi_{a^\ast}(X_t)e^{-a^\ast J_t}\label{eq:Mt}
\end{equation}  
is a martingale,
where $\phi_a(x)$ is the right eigenvector of  the Perron root $\lambda_J(a)$ of the tilted matrix $\tilde{\mathbf{q}}(a)$.   Note that the effective affinity $a^\ast$ appears as the prefactor of $J_t$ in the exponential martingale (\ref{eq:Mt}), and hence effective affinities also play a role in martingale theory for generic currents.  

\subsection{Derivation  of  the martingale property  of $M_t$}

The martingality of $M_t$ follows from the fact that $\phi_{a^\ast}(x)e^{-a^\ast j}$ is a harmonic function of the generator of the joint process $(X_t,J_t)$.
 
The function
\begin{equation}
f_t(x,j) = \langle \phi_{a^\ast}(X_t)e^{-a^\ast J_t}\vert X_0=x,J_0=j\rangle,  \label{eq:def}
\end{equation}
 solves the Kolmogorov backward equation of the joint process $(X_t,J_t)$, i.e.,
    \begin{eqnarray}
    \partial_t f_t(x,j) &=& 
   \sum_{y\in \mathcal{X}} \mathbf{q}_{xy}  [f_t(y,j+c_{x y}) - f_t(x,j)],
    \end{eqnarray}
    with initial condition 
    \begin{equation}
    f_0(x,j) = \phi_{a^\ast}(x)e^{-a^\ast j}.   \label{eq:f0}
    \end{equation}
The process $M_t$ is a martingale if   $\partial_t f_t(x,j) = 0$, see~Ref.~\cite{roldan2022martingales}, which requires that $f_0(x,j)$ is a harmonic function of the joint process $(X,J)$, i.e.,
\begin{equation}
 \sum_{y\in \mathcal{X}}\mathbf{q}_{xy}  [ f_0(y,j+c_{xy}) - f_0(x,j)] = 0. \label{eq:harmonic}
\end{equation} 
Using   (\ref{eq:f0}), we find that (\ref{eq:harmonic}) holds when
\begin{equation}
 \sum_{y\in \mathcal{X}}\tilde{\mathbf{q}}_{xy}(a^\ast)  \phi_{a^\ast}(y)   =   0,
\end{equation}  
which implies that $\phi_{a^\ast}(x)$ is a right eigenvector of the tilted matrix $\tilde{\mathbf{q}}_{xy}(a^\ast)$ associated with a zero root.  Hence $\phi_a^\ast(x)e^{-a^\ast j}$ is a harmonic function and $M_t$ is a martingale.  

Alternatively, we  can use  Radon-Nikodym derivatives, $Q[X^t_0]/P[X^t_0]$ to show that $M_t$ is a martingale~\cite{roldan2022martingales, chetrite_nonequilibrium_2015}.   Specifically,  let $Q[X^t_0]$ denote the probability density of  $X^t_0$ when trajectories are generated by the Markov jump process with rate matrix  $\boldsymbol{\phi^{-1}_{a^\ast}} \tilde{\mathbf{q}}(a^\ast)\boldsymbol{\phi_{a^\ast}}$, where  $\boldsymbol{\phi_{a^\ast}}$ is a diagonal matrix with diagonal entries $\phi_{a^\ast}(x)$.    Then it holds that~\cite{chetrite_nonequilibrium_2015} 
\begin{equation}
M_t = \phi_{a^\ast}(X_0) \frac{Q[X^t_0]}{P[X^t_0]},
\end{equation}
where $P[X^t_0]$ is  the probability density of $X^t_0$ under the dynamics generated by $\mathbf{q}$.   As $M_t$ is a Radon-Nikodym derivative process,  up to a time-independent prefactor, we conclude that   $M_t$ is a martingale~\cite{roldan2022martingales}.

\subsection{Special cases of $M_t$}\label{sec:martingale_special} 

 Equation \eqref{eq:Mt} describes a family of martingales for  currents $J_t$.   For specific choices of the currents $J_t$ we recover known martingales in Markov processes~\cite{roldan2022martingales}.    Notably, for $J_t=S_t$ we get $M_t= \exp(-S_t)$, as  $a^\ast=1$ and $\phi_{a^\ast}=1$, and thus we recover that the exponentiated  negative entropy production  is a martingale~\cite{chetrite2011two, neri2017statistics}.  Another limiting case  is when $J_t=J^{x\rightarrow y}_t$, the edge current between states $x,y\in \mathcal{X}$ (the edge current can be obtained from Eq. \eqref{eq:genericCurrent} by setting $c_{ab} = \delta_{a,x}\delta_{b,y} - \delta_{a,y}\delta_{b,x}$, where $\delta_{a,b}$ is the Kronecker delta function).  In this case, the 
martingale $M_t$, given by Eq.~(\ref{eq:Mt}), is equivalent to the martingale described by Eq.~(69) of  Ref.~\cite{neri2023extreme}, as we show in Appendix \ref{subsec:edgecurrentmartingale}.  This follows from the fact that for edge currents,  $a^\ast$ is given by Eq.~\eqref{eq:effectEdge}, and from an explicit expression for   the right null eigenvector $\phi_{a^\ast}$  that we derive in Appendix~\ref{subsec:edgecurrentmartingale}.
 
\section{First passage problems and extreme value statistics of currents}\label{sec:fpp}

We  revisit  the first passage problem of  a fluctuating current $J_t$ exiting an open interval $(-\ell_-,\ell_+)$ as studied in Ref.~\cite{neri2022universal}.  The first passage time is defined by 
\begin{equation}
T := {\rm min}\left\{t\geq 0: J_t \notin (-\ell_-,\ell_+)\right\}.  \label{eq:FPProb}
\end{equation} 
This is a generalisation of  the gambler's ruin problem, as  introduced by Pascal in the 17th century~\cite{devlin2010unfinished, Huygens}, that  applies to  fluctuating currents~\cite{avanzini2024methods}.   

The splitting probability $p_-$, corresponding with the probability of ruin in the gambler's ruin problem, is the probability that the current first exits the interval $(-\ell_-,\ell_+)$ from the negative threshold, namely, 
\begin{equation}
p_- :=  P\left(J_T\leq -\ell_-\right),
\end{equation}
where the symbol $P$ denotes the probability of an event.  

Using the martingale $M_t$, we show in Sec.~\ref{sec:split} that  the effective affinity is the  decay  constant determining the exponential decay of $p_-$ with $\ell_-$, i.e.,
\begin{equation}
\lim_{\ell_- \rightarrow \infty} \frac{|\ln p_-|}{\ell_-} = a^\ast.    \label{eq:split}
\end{equation}  
As a corollary of this result, we find that $a^\ast$ is also the exponential decay constant of the distribution of current infima $J_{\rm inf}$, see Eq.~(\ref{eq:inf}), relating large deviation theory extreme value statistics.  Next, in Sec.~\ref{sec:fpr} we use martingales to derive the inequality in Ref.~\cite{neri2022universal} between the rate of dissipation, the mean-first passage time, the splitting probability, and the thresholds.   Although this inequality was presented before, the derivation we present here is new and arguably more transparent (it does not involve scaling arguments as in \cite{neri2022universal}).  

\subsection{Splitting probability}\label{sec:split}

Since $\langle J_t\rangle>0$,  and since the interval $(-\ell_-,\ell_+)$ is finite, the first-passage time $T$ is with probability one finite, which implies that
\begin{equation}
p_- + p_+ = 1, \label{eq:1}
\end{equation}
where $p_+ = P(J_T \geq \ell_+)$ and $p_- = P(J_T\leq -\ell_-)$ are the probabilities  that the current  $J_t$ exits the interval $(-\ell_-,\ell_+)$ from the negative or positive thresholds, respectively.  

According to  Doob's Optional Stopping Theorem, see Refs.~\cite{williams1991probability,neri2019integral}, a martingale $M_t$ has the following property
\begin{equation}
\langle M_T\rangle = \langle M_0\rangle,  
\end{equation}
if $T$ is with probability one finite and $|M_T|\leq \kappa$ where $\kappa$ is a constant; notice that both conditions apply as the set $\mathcal{X}$ is finite and $\vert J_T\vert\leq {\rm max}\{\ell_+,\ell_-\} + {\rm max}_{(x,y)\in \mathcal{X}^2}c_{xy}$.     Substitution of  Eq. \eqref{eq:Mt} in Doob's Optional Stopping Theorem yields 
\begin{eqnarray}
\lefteqn{\langle \phi_{a^\ast}(X_T)\exp(-a^\ast J_T) \rangle }&& \nonumber\\ 
&=& p_-  e^{a^\ast \ell_- (1+o_{\ell_{\rm min}}(1))}\langle \phi_{a^\ast}(X_T) \rangle_- +p_+ e^{-a^\ast \ell_+(1+o_{\ell_{\rm min}}(1))}\langle \phi_{a^\ast}(X_T) \rangle_+ 
\nonumber\\ 
&=& \langle \phi_{a^\ast}(X_0)\rangle = 1,\label{eq:2} 
\end{eqnarray} 
where $\langle \cdot \rangle_+$ and $\langle \cdot \rangle_-$ are expectations over repeated realisations of the process $X_t$ conditioned on the event that $J_T\geq \ell_+$ or $J_T\leq -\ell_-$, respectively, and where
 $o_{\ell_{\rm min}}(1)$ represents an arbitrary function that decays to zero when $\ell_{\rm min} = {\rm min}\left\{\ell_-,\ell_+ \right\}$ diverges.     We have used $J_T = \pm\ell_{\pm} (1+ o_{\ell_{\rm min}}(1))$, as the increments of $J_t$ are bounded and independent of $\ell_-$ and $\ell_+$.   In addition, note that, 
without loss of generality, we have  set $\langle \phi_{a^\ast}(X_0)\rangle = 1$, as $\phi_{a^\ast}(x)$ is a right eigenvector of $\tilde{\mathbf{q}}(a^\ast)$, and it is thus defined up to an arbitrary constant.   Solving Eqs.~(\ref{eq:1}) and (\ref{eq:2}) towards  $p_-$ yields\begin{equation}
p_- = \frac{1-e^{-a^\ast \ell_+(1+o_{\ell_{\rm min}}(1))}\langle \phi_{a^\ast}(X_T) \rangle_+}{e^{a^\ast \ell_- (1+o_{\ell_{\rm min}}(1))}\langle \phi_{a^\ast}(X_T) \rangle_-  - e^{-a^\ast \ell_+(1+o_{\ell_{\rm min}}(1))}\langle \phi_{a^\ast}(X_T) \rangle_+ }.
\end{equation}
Taking the limits $\ell_-,\ell_+\rightarrow \infty$, $\ln \langle \phi_{a^\ast}(X_T)\rangle_{\pm}$ can be contained in the correction term $o_{\ell_{\rm min}}(1)$, as the set $\mathcal{X}$ is finite and $|\phi_{a^\ast}(x)|$ is bounded and independent of the thresholds $\ell_-$ and $\ell_+$ and this yields us the formula (\ref{eq:split}) that we were meant to derive.

In the limit of $\ell_+\rightarrow \infty$, the splitting probability $p_-$ is the cumulative distribution of $J_{\rm inf}$,  i.e., 
\begin{equation}
    P\left(J_{\rm inf} \leq \ell_-\right) = \lim_{\ell_+ \rightarrow \infty} p_-(\ell_-,\ell_+) , \label{eq:infRel}
\end{equation}
and Eq.~(\ref{eq:split}) combined with Eq.~(\ref{eq:infRel})  yields the result (\ref{eq:inf}) for the infimum statistics of $J$.

\subsection{Thermodynamic trade off relation  in first-passage setups } \label{sec:fpr}
The inequality (\ref{eq:lowerbound}) combined with  the martingale result  (\ref{eq:split}) implies a trade off relation between dissipation ($\dot{s}$), speed ($\langle T\rangle$), and uncertainty ($|\ln p_-|$)~\cite{neri2022universal}. Indeed, using  Eq.~(\ref{eq:split}) and an asymptotic version of  Wald's equation that applies to  fluctuating currents (see Appendix~\ref{app:Wald} for a derivation)~\cite{Wald1, Wald2, gingrich2017fundamenta}, namely
\begin{equation}
\overline{j} = \frac{\ell_+}{\langle T\rangle}(1+o_{\ell_{\rm min}}(1)), \label{eq:Wald}
\end{equation}
in the inequality~(\ref{eq:lowerbound}), yields 
\begin{equation}
\dot{s} \geq \frac{\ell_+}{\ell_-} \frac{|\ln p_-|}{\langle T\rangle} \left(1+o_{\ell_{\rm min}}(1)\right).   \label{eq:ineq2}
\end{equation}

The trade off inequality (\ref{eq:ineq2}) applies to physical processes with a finite termination time.   An example is a molecular motor that moves on a one-dimensional substrate and detaches from this substrate once it reaches one of its two end points, which we label by the $+$ and $-$ end points.    Assuming that the  motor is biased towards the $+$ end point,  $p_-$ quantifies the probability of reaching the wrong end point, $\langle T\rangle$ is the average duration of the process, and $\dot{s}$ is the rate of dissipation.    Hence, in this example Eq.~(\ref{eq:ineq2}) expresses a trade off between the error, the speed, and the rate of dissipation.   
 
Replacing the quantities $\langle T\rangle$ and $p_-$  in the right-hand side of Eq.~(\ref{eq:ineq2})  with their empirical estimates, we obtain an estimator for $\dot{s}$.    The bias of this estimator is given by $\hat{s}_{\rm FPR}-\dot{s}$, with  
\begin{equation}
\hat{s}_{\rm FPR} := \frac{\ell_+}{\ell_-} \frac{|\ln p_-|}{\langle T\rangle}    \label{eq:SFPR}
\end{equation}
the first-passage ratio~\cite{neri2022estimating}, which gives the average of the estimator.
  In the previous work~\cite{neri2022estimating}, the first-passage ratio was obtained by numerically solving recursion relations for $p_-$ at finite thresholds $\ell_-$ and $\ell_+$.  Instead, here we have shown that 
\begin{equation}
\hat{s}_{\rm FPR} = \overline{j}a^\ast,
\end{equation}
and hence the bias   in the   estimator  can  readily be obtained from diagonalising the tilted matrix $\tilde{\mathbf{q}}(a)$.

The trade off inequality (\ref{eq:ineq2}) is reminiscent of the thermodynamic uncertainty relations~\cite{barato2015thermodynamic,gingrich2016dissipation,gingrich2017fundamenta}, albeit with the important difference that  in (\ref{eq:ineq2})  the uncertainty in the current is quantified with the splitting probability $p_-$, while in the thermodynamic uncertainty relations the uncertainty in the current  is quantified with the variance of $J_t$~\cite{barato2015thermodynamic,gingrich2016dissipation} or the variance of $T$~\cite{gingrich2017fundamenta}. In fact the inequalities (\ref{eq:ineq2}) and (\ref{eq:lowerbound}) are equivalent with the thermodynamic uncertainty relations when  the probability distribution of $J_t$ converges asymptotically with time  to a  Gaussian distribution.  Indeed, it holds then that 
\begin{equation}
    \lambda_J(a) = a(a\sigma^2/2-\overline{j}),
\end{equation}
where $\sigma$ is the standard deviation of $J_t/t$, and thus  
\begin{equation}
a^\ast = 2\overline{j}/\sigma^2.
\end{equation}
Substituting this value into (\ref{eq:lowerbound}) yields the thermodynamic uncertainty relation $\dot{s}\geq 2(\overline{j}/\sigma)^2$~\cite{barato2015thermodynamic, gingrich2016dissipation}.  
  
\section{ Optimal currents, cycle equivalence classes, and relation between effective and thermodynamic affinities}\label{sec:tightness} 
We say that a current $J_t$ is {\it optimal} if the equalities in the bounds \eqref{eq:lowerbound} and \eqref{eq:ineq2} are attained.    For optimal currents, the effective affinity captures all the dissipation in the process $X_t$, and hence  estimates of dissipation based on the first-passage ratio $\hat{s}_{\rm FPR}$ are unbiased.  Optimal currents are also relevant for systems that want to optimise the trade off between dissipation, speed, and uncertainty, as expressed in Eq.~(\ref{eq:ineq2}).   

The stochastic entropy production $S_t$ is an example of an optimal current, as in Sec.~\ref{subsec:EALimiting} we have shown that 
the equality in the bounds \eqref{eq:lowerbound} and   (\ref{eq:ineq2}) are attained for currents of the form $J_t=kS_t$, with $k$ a constant.   In this section, we extend significantly  this result by introducing cycle equivalence classes.  Moreover,  we show that all currents $J_t$ are optimal for  Markov processes defined on unicyclic networks, and therefore in unicyclic networks  effective affinities and thermodynamic affinities are the same.   

This section is structured as follows.
First, in Sec.~\ref{sec:recap},   we partition the set $\mathcal{J}$ of all currents $J_t$  into cycle equivalence classes $[J_t]$.     The cycle equivalence classes are defined so that all currents in the same equivalence class have the same set of cycle coefficients $c_{\gamma}$, with $\gamma$ the \textit{fundamental cycles} that are part of the \textit{fundamental cycle basis} $\mathcal{C} \reflectbox{$\in$}  \gamma$; a fundamental cycle basis is a cycle basis of a graph constructed by adding chords to a spanning tree~\cite{bollobas_modern_1998}. 
In the statistical physics literature  cycle decompoisitions of currents are known from Schnakenberg's network theory~\cite{schnakenberg,baratoSymmetryCurrentProbability2012}, and we review some relevant parts of this theory also in Sec.~\ref{sec:recap}.    Next, in Sec.~\ref{sec:samecyclesame} we show that all currents that are part of the same cycle equivalence class $[J_t]$ have the same effective affinity $a^\ast$.   Therefore, the optimality property of $kS_t$ extends to all currents that belong to the  cycle equivalence classes $[kS_t]$.  Note that equivalence classes  are sets of currents that are isomorphic to $\mathbb{R}^{|\mathcal{X}|-1}$, and all the currents in the sets $[kS_t]$ are optimal in the sense that  the equalities in the  Eqs.~\eqref{eq:lowerbound} and   (\ref{eq:ineq2}) are attained. Lastly, in Sec.~\ref{sec:effeThermod}, we show that for systems fully described by a set of uncoupled currents, such as unicyclic systems, the effective affinities are identical to the thermodynamic affinities.

\subsection{Definition of  cycle equivalence classes  $[J]$}\label{sec:recap}
   
Let $\mathcal{G}=(\mathcal{X},\mathcal{E})$ be the graph of admissible transitions of the Markov jump process  $X$ defined by $\mathbf{q}$. 
 As detailed in  Appendix~\ref{sebsec:fundamentalbasis},  we  can construct a fundamental cycle basis $\mathcal{C}$   that spans  the cycle space of the graph $\mathcal{G}$  (any cycle in the graph of admissible transitions can be obtained as a symmetric difference of the  fundamental cycles $\gamma \in \mathcal{C}$).  
The cycles in  $\mathcal{C}$ are sequences of nonrepeating vertices  $[x_1(\gamma),x_2(\gamma),\ldots, x_{n(\gamma)}(\gamma),x_1(\gamma)]$, where  $n(\gamma)$ is the number of distinct vertices in $\gamma$ and the $x_i$ are the vertices visited in the cycle.   

  Given a fluctuating current $J_t$ that is determined by the coefficients $c_{xy}$ with $(x,y)\in\mathcal{E}$ [see Eq.~(\ref{eq:genericCurrent})],  we define its cycle coefficients $c_{\gamma}$ by 
\begin{equation}
    c_\gamma := \sum_{i=1}^{n(\gamma)} c_{x_i x_{i+1}} ,  \quad \gamma\in \mathcal{C},  \label{eq:cgamma} 
\end{equation} 
where the $x_i$ are the vertices of $\gamma$, and 
 it should be understood  that $x_{n(\gamma)+1}=x_1$.   The cycle coefficients  $c_{\gamma}$  quantify the amount of resource transported when the process $X$ traverses the cycle $\gamma$ once.  

 Next, we define the currents $\overline{j}_{\gamma}$ associated with the fundamental cycles $\gamma\in \mathcal{C}$~\cite{schnakenberg, baratoSymmetryCurrentProbability2012}.   
At stationarity, $\partial_t p_t(x)=0$, and hence (\ref{eq:9}) implies that  
\begin{equation}
    \sum_{y\in \mathcal{X}:(x,y)\in \mathcal{E}} \overline{j}_{xy} = 0,\label{eq:KCL}
\end{equation}
which we recognise as Kirchhoff's first law applied to each state $x\in\mathcal{X}$.  
This set of $|\mathcal{X}|$ equations imposes $|\mathcal{X}| -1$ linearly independent constraints on the edge currents $\overline{j}_{xy}$ (and not $|\mathcal{X}|$ constraints as $\sum_{x\in \mathcal{X}}p_t(x)$ must be constant in time).    Consequently the number of independent currents required to describe all the edge currents is $|\mathcal{C}| = |\mathcal{E}| - |\mathcal{X}|+1$. We choose  $|\mathcal{C}|$ such currents by associating to each fundamental cycle $\gamma \in \mathcal{C}$ a current $\overline{j}_\gamma$ which satisfy the equation
\begin{equation}
    \overline{j}_{xy} = \sum_{\gamma \in \mathcal{C}} \eta_{x,y,\gamma} \overline{j}_{\gamma}, \label{eq:gammadefine}
\end{equation}
where $\eta_{x,y,\gamma}=0$ if the edge $(x,y)$ is not in the cycle $\gamma$, and $\eta_{x,y,\gamma}=+1 (-1)$ if the edge $(x,y)$ is in the cycle and the node $x$ appears before (after) the node $y$ in the cycle $\gamma$. Using the constraints \eqref{eq:KCL}, the set (\ref{eq:gammadefine}) of 
$|\mathcal{E}|$ equations reduces 
to a set of $ |\mathcal{E}| - |\mathcal{X}|+1 = |\mathcal{C}|$ independent equations that can be solved towards the  $|\mathcal{C}|$  cycle currents $\overline{j}_\gamma$.

In summary, for a given Markov process, we choose a fundamental cycle basis $\mathcal{C}$ as described in Appendix.~\ref{sebsec:fundamentalbasis}. We then  associate a cycle coefficient $c_\gamma$ for each $\gamma \in \mathcal{C}$ by Eq.~\eqref{eq:cgamma}. We also associate a cycle current $\overline{j}_{\gamma}$ for each $\gamma \in \mathcal{C}$ by solving the equations \eqref{eq:gammadefine}. Then, using these quantities $\overline{j}_\gamma$ and $c_\gamma$ as defined, and Eq.~\eqref{eq:javgsum}, the average current $\overline{j}$ of any fluctuating current $J_t$ is expressible as
\begin{equation}
    \overline{j} = \sum_{\gamma \in \mathcal{C}} c_{\gamma}\overline{j}_{\gamma}.   
\end{equation}
Therefore the coefficients $c_{\gamma}$ are sufficient to determine the current $\overline{j}$ associated with a fluctuating current $J_t$. For example, for  $J_t = S_t$ as defined in Eq. \eqref{eq:St}, we can write 
\begin{equation}
    \dot{s} = \sum_{\gamma\in\mathcal{C}} a_\gamma j_\gamma
\end{equation}
as expressed before in Eq.\eqref{eq:agamma} where
\begin{equation}
    a_\gamma = \sum_{i=1}^{n(\gamma)} \ln{\frac{k_{x_ix_{i+1}}}{k_{x_{i+1}x_i}}}\label{eq:agamma1}
\end{equation}
and as before $x_i$ are the vertices of $\gamma$ with $x_{n(\gamma)+1}=x_1$. In  Sec.~\ref{sec:samecyclesame}, we will show that the coefficients $c_\gamma$ are also sufficient to determine $\lambda_J(a)$ for a given current $J_t$.

We define cycle equivalence classes as the  set of currents   that have the same  cycle coefficients $c_{\gamma}$, and we write $[J_t]$ to denote  the cycle equivalence class that contains the current $J_t$.    Note that the set $\mathcal{J}$  of all currents is  isomorphic to the Euclidean space $\mathbb{R}^{|\mathcal{E}|}$.  Since there are  a number $|\mathcal{C}| = |\mathcal{E}| - |\mathcal{X}|+1$  of cycle coefficients (see Appendix~\ref{sebsec:fundamentalbasis}), the cycle equivalence classes are isomorphic to $\mathbb{R}^{|\mathcal{X}|-1}$.    In the next section, we show that all fluctuating currents $J_t$ within a cycle equivalence class have the same effective affinity, which is our interest in these equivalence classes.

\subsection{Currents in the same cycle equivalence class have the same effective affinity} \label{sec:samecyclesame}
 We  show  that all currents in the same cycle equivalence class $[J_t]$ have the same  logarithmic moment generating function $\lambda_J(a)$, and hence also the same effective affinity $a^\ast$.  To this aim, we show that  $\lambda_J(a)$ depends on the coefficients $c_{xy}$ through the cycle coefficients~$c_{\gamma}$.    

\begin{figure}
    \centering
    \includegraphics[width=0.6\linewidth]{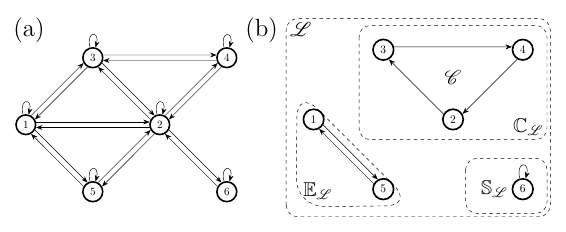}
    \caption{ Example of a  spanning linear subgraph $\mathscr{L}$ as used in the graphical expansion of the characteristic polynomial of $\tilde{\mathbf{q}}$ in Eq.~(\ref{eq:charpol1}).   Panel (a) shows an example of a graph of admissible transitions $\mathcal{G}=(\mathcal{X},\mathcal{E})$ in a Markov chain with  six states. For the graphical expansion we need to represent $\mathcal{G}=(\mathcal{X},\mathcal{E})$   as a directed graph (each nondirected edge equals two directed edges) and we need to associated  to each node a self-loop.    Panel (b) shows a spanning linear subgraph of the graph shown in Panel (a).}
    \label{fig:2}
\end{figure}

Since $\lambda_J(a)$ is the Perron root of $\tilde{\mathbf{q}}(a)$, it is sufficient to demonstrate that the characteristic polynomial of 
 the tilted matrix $\tilde{\mathbf{q}}(a)$ depends on the coefficients $c_{xy}$ through the cycle coefficients~$c_{\gamma}$.      
 To show this latter, we consider a graphical expansion of the characteristic polynomial of $\tilde{\mathbf{q}}(a)$ in terms of the spanning, linear subgraphs $\mathscr{L}$ of the graph of admissible transitions, $\mathcal{G}=(\mathcal{X},\mathcal{E})$, of  the Markov process $X_t$.  
 We provide a formal definition of  $\mathscr{L}$ in the Appendix~\ref{subsec:S5A}, while here we provide an informal definition with the help of   Fig.~\ref{fig:2}.  A linear subgraph of the graph is a directed subgraph for which the indegree and outdegree of each node equals one, and we call the subgraph spanning when its vertex set equals $\mathcal{X}$.  
 To construct the  spanning, linear subgraphs of  $\mathcal{G}=(\mathcal{X},\mathcal{E})$  we use two more conventions, namely, we consider that all non-directed edges of $\mathcal{E}$ consist of two directed edges and we add to all nodes a self-loop  (see panel (a) of Fig.~\ref{fig:2}).  
 In Panel (b) we show an example of a spanning linear subgraph.   It is the disjoint union of   three kind of graphs, namely, self-loops ($\mathbb{S}_\mathscr{L}$) that consist of an isolated node,  double edges ($\mathbb{E}_{\mathscr{L}}$) that consist of two nodes connected by a non-directed edge (or equivalently two directed edges), and directed simple cycles ($\mathbb{C}_{\mathscr{L}}$).  
 
 As we show in Appendix~\ref{sec:S5}, the characteristic polynomial of $\tilde{\mathbf{q}}(a)$ can be expressed as a sum over all spanning, linear subgraphs of the graph of admissible transitions in the Markov process $X$, viz., 
\begin{equation}
\begin{split}
{\rm det}\left( \tilde{\mathbf{q}}(a) - \xi \mathbbm{1} \right)= \sum_{\mathscr{L}}  (-1)^{\vert\mathcal{X}\vert+\kappa(\mathscr{L})} \left(\prod_{(x,x)\in\mathbb{S}_{\mathscr{L}}}\left(\mathbf{q}_{xx} - \xi\right)\right)\left(\prod_{{\left\{(x\rightarrow y),(y\rightarrow x)\right\}\in\mathbb{E}_\mathscr{L}}}\mathbf{{q}}_{yx}\mathbf{{q}}_{xy} \right)  \\\left(\prod_{\mathscr{C}\in\mathbb{C}_{\mathscr{L}}} \exp\left(\mathcal{A}_\mathscr{C}\right)\exp\left(-a\sum_{\gamma \in \mathcal{C}} \epsilon_{\mathscr{C},\gamma} c_\gamma\right) \right), 
\label{eq:charpol1}
\end{split} 
\end{equation} 
where $ {\rm det}(\cdot)$ denotes the determinant of a matrix,  
$\kappa(\mathscr{L})$  is the number of strongly connected components in the spanning linear subgraph $\mathscr{L}$, $\mathcal{A}_\mathscr{C} = \ln \prod_{(x\rightarrow y)\in\mathscr{C}} \mathbf{q}_{yx} $, and $\mathbbm{1}$ is the identity matrix of order $|\mathcal{X}|$.   The Eq.~(\ref{eq:charpol1})  follows from  the Coefficients Theorem for Directed Graphs~\cite{acharya_spectral_1980}, see Appendix~\ref{sec:S5}.   An expression similar to (\ref{eq:charpol1})  for the   characteristic polynomial of $\tilde{\mathbf{q}}(a)$  was derived in Ref.~\cite{baratoSymmetryCurrentProbability2012} in the context of the Gallavotti-Cohen fluctuation symmetry.  

It readily follows from Eq.~(\ref{eq:charpol1}) that the dependency of the characteristic polynomial of $\tilde{\mathbf{q}}(a)$  on the coefficients $c_{xy}$ is through the cycle coefficients $c_{\gamma}$, and therefore all currents in $[J_t]$ share the logarithmic moment generating function $\lambda_J(a)$ and the same effective affinity.   

In addition, in Appendix  \ref{subsec:S5B} we show  that all currents in $[kS_t]$ satisfy the Galavotti-Cohen fluctuation symmetry \eqref{eq:galavottiCohen}, which is a sufficient condition for the equality in \eqref{eq:lowerbound} to be achieved (as shown in Sec.\ref{subsec:EALimiting}). Thus all currents in the equivalence classes $[kS_t]$ are optimal. 

\subsection{Equality of the effective and thermodynamic affinities for  systems with uncoupled currents}\label{sec:effeThermod}
We show that for systems fully described by uncoupled currents, the thermodynamic affinity of a current equals its effective affinity.   

Let us first consider the case of unicyclic Markov processes with $\vert\mathcal{C}\vert=1$.  
A direct consequence of the optimality of all currents in the classes $[kS_t]$ is that for unicyclic Markov processes all currents $J_t$ are optimal, and thus 
    $\dot{s} = a^\ast \overline{j}$.  By comparison with Eq.~(\ref{eq:sdotMacro}) for one current, we conclude that $a^\ast$ is the thermodynamic affinity.   

For systems that are fully described by a set of $n$ uncoupled currents, the graph of admissible transitions  consists of $n$ connected components, each of which is unicyclic. Thus $\vert\mathcal{C}\vert=n$. Associating a fluctuating current $J^\gamma_t$ to each cycle $\gamma \in \mathcal{C}$, we find 
\begin{equation}
    \dot{s}_\gamma = a^\ast_\gamma \overline{j}_\gamma,
\end{equation}
where $\dot{s}_\gamma$ is the rate of dissipation  corresponding with the connected component that contains $\gamma$, $a^\ast_\gamma$ is the effective affinity of $J^\gamma_t$, and $\overline{j}_\gamma$ is the average current associate with $J^\gamma_t$.  Thus, the total  rate dissipation of the system is given by 
\begin{equation}
   \dot{s} = \sum_{\gamma\in\mathcal{C}} a^\ast_\gamma \overline{j}_\gamma.
\end{equation}
Comparing the above equation with Eq.~\eqref{eq:sdotMacro},  we find that for systems fully described by uncoupled currents, the effective affinity of each uncoupled current is the same as its corresponding thermodynamic affinity. 

\section{Necessary condition for optimal currents in a toy model with two cycles}\label{sec:twocycle} 
Equivalence with the stochastic entropy production $S_t$ is, up to an irrelevant proportionality constant, sufficient for optimality of a current (i.e., so that the equalities in Eqs. \eqref{eq:lowerbound} and ~\eqref{eq:ineq2} are attained).   However, the question remains whether this condition is necessary, i.e., whether the cycle equivalence classes  $[kS_t]$ contain all possible currents that attain the equalities in the bounds (\ref{eq:lowerbound})  and (\ref{eq:ineq2}).  

\begin{figure}
\centering

\tabskip=0pt

\includegraphics{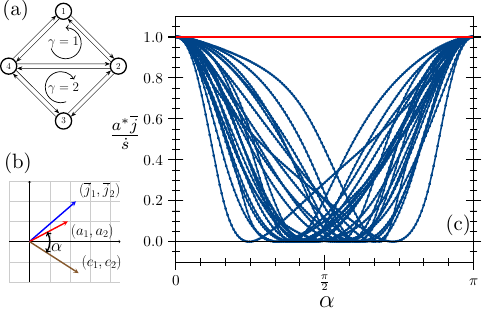}

    \caption{Panel(a): Graph of admissible transitions for the  four state model with the two cycles $\gamma=1$ and $\gamma=2$ as indicated.  Panel (b): Sketch of cycle affinities $(a_1,a_2)$, cycle currents $(j_1,j_2)$, and cycle coefficients $(c_1,c_2)$ plotted in $\mathbb{R}^2$, with the angle $\alpha$ indicated.     Panel (c):  $a^\ast \overline{j}/\dot{s}$ as a function of  $\alpha$ for $\overline{j}=1$.  Different lines correspond to different choices of  the rates  $\mathbf{q}_{xy}$, here randomly generated with uniform distribution between $0$ and $1$. 
} 
    \label{fig:3}
\end{figure}

 Here, we settle this question for models with two fundamental cycles through  
 a numerical case study of the  four state model illustrated in Fig.~\ref{fig:3}(a).       The four state model  has  two fundamental cycles denoted by $\gamma=1$ and $\gamma=2$, and hence the cycle equivalence classes of this model are determined by two coefficients $c_1$ and $c_2$, such that  $\overline{j} = c_1\overline{j}_1 + c_2\overline{j}_2$.    
We normalise  $c_1$ and $c_2$ such that $\overline{j}=1$.    For this choice of normalisation, the dependence of $a^\ast$ on the $c_{xy}$-coefficients that define $J_t$ is fully determined by one parameter, namely the angle $\alpha$ between the vectors $(c_1,c_2)$ and $(a_1,a_2)$ in $\mathbb{R}^2$, where the latter are the cycle coefficients of  $[S_t/\dot{s}]$; see  Fig.~\ref{fig:3}(b) for an illustration.   

Figure~\ref{fig:3}(c) plots  $a^\ast$ as a function of $\alpha$ for randomly generated transition rates $\mathbf{q}$.   According to the inequality (\ref{eq:lowerbound}), which here reads $a^\ast/\dot{s}\leq 1$,    the equality $a^\ast= \dot{s}$ is attained when $\alpha=0$ or $\alpha=\pi$, corresponding with fluctuating currents that belong to  $[S_t/\dot{s}]$ or $[-S_t/\dot{s}]$, respectively.   We observe that the effective affinity is a monotonously decreasing/increasing function between the value of $\alpha$ with vanishing average current (where $a^\ast=0$) and the end point values  $\alpha=0$ and $\alpha=\pi$. 
Hence, for models with two fundamental cycles, the equalities in the trade-off relations (\ref{eq:lowerbound}) and (\ref{eq:ineq2}) are  only attained for currents that belong to the cycle equivalence classes $[kS_t]$ with $k\in\mathbb{R}$.      

In conclusion, for models with two cycles the  equivalence of currents with $S_t$ (in the sense of cycle equivalence classes) is necessary and sufficient  for optimality.

\section{Effective affinity for a driven Brownian particle on a ring}\label{sec:diffusion}

Thus far, we have focused  on Markov  processes defined on a discrete set $\mathcal{X}$.   Nevertheless,  the definition of the effective affinity, Eq.~(\ref{eq:defa}), and its ensuing properties --- amongst others, the bound (\ref{eq:lowerbound}), the infimum statistics results (\ref{eq:inf}), and the martingale $M_t$ ~(\ref{eq:Mt}) ---   also apply to driven diffusions.  

As an illustration of the applicability of the effective affinity to driven diffusions,   we  analyse in this section the effective affinity for  a particle that is subject to a  constant non-conservative force $f$ and undergoes overdamped diffusion on a ring.  We show that in this model the  effective affinity of an arbitrary current   equals the macroscopic (thermodynamic) affinity.  Furthermore,  we derive an explicit expression for the martingale $M_t$, and we show that $M_t$ is different from the exponentiated negative entropy production.   Lastly, we show that the equality in Eq.~(\ref{eq:lowerbound}) is attained.   These properties are similar to those of currents in Markov jump processes defined on unicyclic graphs.    

\subsection{Currents in a Brownian particle on a ring}
We consider a particle that is bound to a  one-dimensional ring and is immersed into an isothermal environment.  The evolution in time of the  particle's position, $X_t \in [0,1)$, is described by the stochastic differential equation 
\begin{equation}
\frac{{\rm d}X_t}{{\rm d}t} = \frac{f}{\gamma} + \sqrt{2D} \frac{{\rm d}W_t}{{\rm d}t}, \label{eq:brownianDif}
\end{equation}
where $f$ is the non-conservative force acting on the particle,  $\gamma$ is the frictional coefficient,  $D$ is the particle's diffusion coefficient, and   $W_t$ is a standard Wiener process representing the stochastic force that  the environment exerts on the particle.  We assume that the frictional coefficient is related to  temperature by Einstein's relation 
\begin{equation}
D = \frac{k_{\rm_B}\mathsf{T}_{\rm env}}{\gamma} \label{eq:einstein}
\end{equation}
where $\mathsf{T}_{\rm env}$ is the temperature.

 The probability density function $p_t(x)$ of the particle's position evolves according to the Fokker-Planck equation 
\begin{equation}
    \partial_t p_t(x)  = \mathcal{L}^\dagger p_t(x),
\end{equation}
with 
\begin{equation}
    \mathcal{L}^\dagger= -f\partial_x  + D \partial^2_x,
\end{equation}
and with periodic boundary conditions, $p_t(x+ 1) =p_t(x)$.

For one-dimensional driven diffusions,   fluctuating currents $J_t$ are Stratonovich integrals of the form~\cite{Seki}
\begin{equation}
    J_t = \int^{t}_{0} c(X_s) \circ {\rm d}X_s,  \label{eq:strato}
\end{equation}
where $c(x)$ is  a bounded, real-valued function on the interval $[0,1)$.  
The Stratonovich   convention, denoted by $\circ$,   ensures that $J_t$ is antisymmetric under the time reversal.  In the example of  $c(x) = 1$ it holds that $J_t$ is the net distance travelled by the particle along the ring.    

In what follows, we assume, without loss of generality, that $f>0$ and  we  set   $\int^{1}_{0} c(x) {\rm d}x = 1$.  In this case, the average current is given by 
\begin{equation}
\overline{j} = \frac{f}{\gamma}. \label{eq:JO}
\end{equation}

\subsection{Effective affinity} 
Just as was the case for Markov jump processes, we define the effective affinity as the nonzero root of the logarithmic moment generating function, i.e., through Eq.~(\ref{eq:defa}), where $\lambda_J$ is defined as in Eq.~(\ref{eq:deflambda}) with the averages $\langle \cdot \rangle$  now over different realisations of the diffusion process Eq.~(\ref{eq:brownianDif}).

We show that for  an  arbitrary current of the form (\ref{eq:strato}),  determined by the function $c(x)$,  the effective affinity takes the form 
\begin{equation}
    a^\ast = \frac{f}{k_{\rm_B}\mathsf{T}_{\rm env}}, \label{eq:EAOver}
\end{equation}
and hence in this model it equals the macroscopic (thermodynamic) affinity divided by the temperature, and the effective affinity is also the stalling force divided by the temperature.   

The tilted operator that governs the evolution of $\langle \exp(-aJ_t) \rangle$ reads~\cite{touchette2009large,tsobgni2016large}
\begin{equation}
    \tilde{\mathcal{L}}(a) = \frac{f}{\gamma}(\partial_x - a c(x)) + D(\partial_x -a c(x))^2,
\end{equation}
and, as for Markov jump processes,  the Perron root of   $    \tilde{\mathcal{L}}(a)$ equals $\lambda_J(a)$. 
Hence,   to find the effective affinity $a^\ast$ we  solve the eigenvalue problem
\begin{equation}
    \tilde{\mathcal{L}}(a^\ast)[\phi_{a^\ast}(x)] = \lambda_J(a^\ast) \phi_{a^\ast}(x) = 0 \label{eq:tiltedL}
\end{equation}
with the periodic boundary condition $\phi_{a^\ast}{(x+1)} = \phi_{a^\ast}{(x)}$. Note that  we are solving (\ref{eq:tiltedL}) towards both $a^\ast$ and $\phi_{a^\ast}$, the latter being the right eigenvector of $\tilde{\mathcal{L}}(a)$ with eigenvalue $\lambda_J(a)$, and $a^\ast$ the value of the parameter $a$ for which the $\lambda_J(a)$ vanishes.

Making the transformation~\cite{proesmans_large-deviation_2019}
\begin{equation}
    \phi_a(x)   = \exp\left(a\int^{x}_{0} c(y) {\rm d}y\right)\psi_a(x),
\end{equation}
we find that Eq.~(\ref{eq:tiltedL}) is equivalent to  the differential equation
\begin{equation}
    \frac{f}{\gamma}\partial_x\psi_{a^\ast}(x) + D \partial_x^2 \psi_{a^\ast}(x) = 0,\label{eq:diffeq}
\end{equation}
which is subject to the boundary condition
\begin{equation}
    \psi_{a^\ast}(x+1) = \psi_{a^\ast}(x) \exp\left( -{a^\ast} \int^{1}_{0} c(y){\rm d}y\right) = {\rm e}^{-{a^\ast}} \psi_{a^\ast}(x),\label{eq:boundcond}
\end{equation}
and we have used here that $\int^{1}_{0} c(y){\rm d}y=1$.  
The equation \eqref{eq:diffeq} admits the general solution
\begin{equation}
  \psi_{a^\ast}(x) = \zeta_1 \frac{ D \gamma}{f} \exp\left(-\frac{f }{D \gamma}x\right) + \zeta_2.
\end{equation}
To satisfy the boundary conditions  \eqref{eq:boundcond}, we set  $\zeta_1 = 1$ and $\zeta_2 =0$, yielding 
\begin{equation}
\psi_{a^\ast}(x)= \frac{D \gamma }{f} \exp\left(-\frac{f }{D \gamma}x\right). \label{eq:sast}
\end{equation}
Substituting Eq.~(\ref{eq:sast}) into (\ref{eq:boundcond}), and using Einstein's relation (\ref{eq:einstein}),  we find the result  Eq.~(\ref{eq:EAOver})  for the effective affinity that we were meant to derive.  

\subsection{Martingale}
In this  example, the martingale Eq.~(\ref{eq:Mt})  reads
\begin{equation}
M_t = \frac{k_{\rm_B}\mathsf{T}_{\rm env}}{f} \exp\left( -\frac{f}{k_{\rm_B}\mathsf{T}_{\rm env}}  \left(\int^{t}_0 c(X_s) \circ {\rm d}X_s  - \int^{X_t}_0 (c(y)-1){\rm d}y \right) \right) ,
\end{equation}
where the second term $ - \int^{X_t}_0 (c(y)-1){\rm d}y $ is a contribution from  the right eigenvector $\phi_{a^\ast}(X_t)$; notice that the second term is not time extensive as $\int^1_0c(x){\rm d}x=1$ .   
In the specific case when $c = 1$, we get 
\begin{equation}
M_t =  \frac{k_{\rm_B}\mathsf{T}_{\rm env}}{f}  \exp\left(-\frac{f}{k_{\rm_B}\mathsf{T}_{\rm env}}X_t\right) =  \frac{k_{\rm_B}\mathsf{T}_{\rm env}}{f} \exp\left(-S_t\right),
\end{equation}
with $S_t = -fX_t/(k_{\rm_B}\mathsf{T}_{\rm env})$ is  the stochastic entropy production~\cite{Seki}, 
and we thus recover the martingale of Ref.~\cite{generic}.  

\subsection{Dissipation bound: all currents are optimal}
We show that  for all currents $J$ the equality in Eq.~(\ref{eq:lowerbound}) is attained, and hence all currents are optimal.

Indeed, since the rate of dissipation equals (see e.g., Chapter 5 of Ref.~\cite{roldan2022martingales})
\begin{equation}
\dot{s} = \frac{f^2}{\gamma k_{\rm_B}\mathsf{T}_{\rm env}}  ,   \label{eq:SDotBrown}
\end{equation}
it follows from Eqs.~(\ref{eq:EAOver}), (\ref{eq:JO}), and (\ref{eq:SDotBrown}) that
\begin{equation}
a^\ast \overline{j} = \dot{s},
\end{equation}
and therefore the equality in (\ref{eq:lowerbound}) is attained.  

The optimality of currents in the present model can also be understood from an application of cycle equivalence classes to a discretised version of the model.   Indeed, since the discretised model is unicyclic, the graph of admissible transitions has one fundamental cycle, and hence all  currents belong to the same 
cycle equivalence class (up to a constant).   Note that since this argument is based on the topology of the  graph of admissible transitions,  we predict that the equalities in Eqs.~(\ref{eq:lowerbound}) and (\ref{eq:ineq2}) are also attained for diffusions on a ring with a position-dependent  force~\cite{tsobgni2016large}.

\section{Effective affinity and stalling forces in a Kinesin-1 model}\label{sec:kinesin} 
So far, we have shown that the effective affinity is closely related to the dissipation of a current and to its fluctuations.   Nevertheless, the question remains  whether we can give a  mechanical interpretation to the effective affinity, in the sense of a  generalised force.  Since the current, $\overline{j}$, vanishes when the effective affinity equals zero, $a^\ast$ cannot represent the thermodynamic force acting on the system, yet it may still be  the current's stalling force (as the stalling force must be $0$ when $\overline{j}=0$).  In this section, we make a study of the potential  connection between the effective affinity and stalling forces.     

First in Sec.~\ref{sec:stallForce} we review results for stalling forces from Ref.~\cite{polettini2019effective}, where the problem was solved for edge currents.  Next, in Sec.~\ref{sec:biophysKin1} we consider a case study on a biophysical model for Kinesin-1 for which the positional current is not an edge current, and thus the generic effective affinity studied in this paper is relevant.   In Sec.~\ref{sec:EASta} we show that  in this biophysical  model the effective affinity is approximately equal to the stalling force, and we show that this property is lost when a set of thermodynamic consistency conditions are violated.   These findings suggest that  effective affinities and stalling forces are related, but  delineating the precise relationship between these two quantities beyond the present toy model remains an  open problem.

\subsection{Stalling forces}\label{sec:stallForce}

Markov processes modelling physical systems often depend on external parameters such as chemical potentials or physical forces, which modify the rates of the Markov process. An interesting property of the edge current effective affinity \eqref{eq:effectEdge}  is that it has a physical interpretation in terms of these external  parameters. 

Let the current of interest be  $J^{xy}_t$, i.e., the net number of jumps along the edge $(x,y)$, and let us assume that a physical force $f$ acts on the system through the $(x,y)$ edge by modifying the rates $\mathbf{q}_{xy}$ and $\mathbf{q}_{yx}$.   An example of such a conjugate current-force pair  is discussed in Refs.~\cite{Liepelt,neri2023extreme}  where $J^{xy}_t$  is the positional current of a molecular motor and $f$ is the mechanical force acting on the motor.  Using the principle of local detailed balance~\cite{maes2021local, peliti2021stochastic},  we can express the ratio of the rates by  
 \begin{equation}
 \frac{\mathbf{q}_{xy}}{\mathbf{q}_{yx}} = \frac{\mathbf{q}^{(0)}_{xy}}{\mathbf{q}^{(0)}_{yx}}\exp\left(-\frac{f d}{k_{\rm_B}\mathsf{T}_{\rm env}}\right),
 \end{equation}
  where the minus sign indicates that the force opposes the positive direction  of flow in the definition of $J^{xy}_t$,  where $\mathsf{T}_{\rm env}$ is the temperature, $k_{\rm_B}$ is the Boltzmann constant, and where  $d$ is the distance crossed by the system when $X$ jumps from $x$ to $y$.    The  matrix $\mathbf{q}^{(0)}$ denotes the transition rates when the  force $f=0$.

 As shown in Ref.~\cite{polletini1}, the effective affinity equals the \textit{additional} force  required to stall the current (i.e, to ensure $\overline{j}^{xy}=0$), which we call the stalling force.  Hence, the effective affinity takes the form  
\begin{equation}
    a^\ast = \frac{(f_0 - f)d}{k_{\rm_B}\mathsf{T}_{\rm env}}=: \mathfrak{f}_{\rm stall}
, \label{eq:toverify}
\end{equation}
where $f_0$ is the stalling force when $f=0$ and $f_0-f$ is the stalling force for nonzero $f$.    Notice that for convenience we introduce the symbol $\mathfrak{f}_{\rm stall}$ for the dimensionless form of the stalling force.   

For edge currents, $f_0$  is defined by the condition that for $f=f_0$ it holds that  
\begin{equation}
 r_{\rm ss}(x)\mathbf{r}_{xy} -r_{\rm ss}(y)\mathbf{r}_{yx}  =0, 
\end{equation}
where  $\mathbf{r}$ is the matrix of transition rates
\begin{equation}
\mathbf{r}_{x'y'} = \left\{\begin{array}{ccc} \mathbf{q}^{(0)}_{x'y'} &{\rm if}& (x'y')\neq (xy), \\\mathbf{q}^{(0)}_{xy} \exp\left(-\frac{f_0d}{k_{\rm_B}\mathsf{T}_{\rm env}}\right)&{\rm if}& (x'y')= (xy), \end{array}\right.
\end{equation}
and 
where $r_{\rm ss}$
is the stationary probability mass function of the Markov process with rate matrix~$\mathbf{r}$.  

In the next section,  we  verify whether (\ref{eq:toverify})  still holds in   a biophysical model of Kinesin-1 for which the positional current is not an edge current.

\subsection{Biophysical model  for Kinesin-1}\label{sec:biophysKin1}
 We define a biochemical model for the motor protein Kinesin-1~\cite{shen2022mechanochemical} that we use to study effective affinities and stalling forces in motor proteins.   Kinesin-1 proteins are  motor proteins that  bind to quasi one-dimensional biofilaments.  Once they are bound to their substrate, the Kinesin-1 motors  perform directed motion towards the biofilament's plus end.   This directed motion is fueled by a hydrolysis reaction that converts adenosine triphosphate (ATP) into adenosine diphosphate  and an inorganic phosphate molecule.   The free energy released in this  reaction is converted into work that generates the directed runs.   Hence,  the entropy production takes the form 
\begin{equation}
\dot{s}=\frac{\Delta \mu}{k_{\rm_B}\mathsf{T}_{\rm env}} \overline{j}_{\mathrm{[ATP]}}-\frac{f d}{k_{\rm_B}\mathsf{T}_{\rm env}} \bar{j}_{\mathrm{pos}},
\end{equation}
where $\Delta \mu = \mu_{\rm ATP}-\mu_{\rm P}-\mu_{\rm ADP}$ is the chemical potential difference between the reagents and the products of the hydrolysis reaction (ATP$\rightarrow$ ADP+P), $f$ is the mechanical force opposing the motion of kinesin-1 towards the microtubule plus end, $d$ is the distance covered in one motor step,  $\mathsf{T}_{\rm env}$ is the temperature of the environment, and $k_{\rm B}$ is the Boltzmann constant (as before).   The currents  $\overline{j}_{\mathrm{[ATP]}}$ and $\overline{j}_{\mathrm{pos}}$ count, respectively, the number of ATP molecules hydrolysed and the number steps made by the motor in a unit of time.    

Next, we consider a Markov jump process that provides a microscopic model for the dynamics of Kinesin-1 motors, viz., we consider the model of Ref.~\cite{shen2022mechanochemical}.  This biophysical model  of Kinesin-1 consists of five states, each corresponding to specific chemical and physical configurations of the two motor heads of Kinesin-1. In each of these states, each head of the motor can either be attached or detached from the microtubule, and also be in any one of ATP, or ADP$\cdot$Pi binding state or be unbound. The graph of admissible  transitions of this model is shown  in Fig.~\ref{fig:4}(a), and as indicated therein, can be decomposed into three fundamental cycles. 
The two ``forward" cycles indicated in Fig.~\ref{fig:4}(a) correspond to a series of transitions in which the motor takes one full step forward. The other cycle represents a series of transitions where the motor does not move forward, and is hence labeled ``futile". 
Furthermore, each step of the motor consists of two substeps, corresponding to the attachment of the head closer to the plus end of the microtubule (the forward head) and the detachment of the head further away (the rear head). This is consistent with experimental data~\cite{coppin1996detection, nishiyama2001substeps}. 
The distinction between the two ``forward" cycles is chiefly that in cycle $1231$ the rear head of the motor detaches from the microtubule before ATP is bound to the forward head, while in cycle $2342$ the ATP binding on the forward head occurs before the rear head detaches from the microtubule, thus representing distinct pathways for forward stepping.  A detailed explanation of all the states and the transitions in the model may be found in Fig.1 of \cite{shen2022mechanochemical}. 
The positional current $J_{\rm pos}$, which gives the displacement of the molecular motor,  sums up contributions   the edges $(1,3)$, $(4,2)$ and $ (2,3)$, accounting for both the ``forward" cycles (see Appendix~\ref{sec:fig3params} for details).

We discuss the dependency of the rates $\mathbf{q}_{xy}$ in the model  on the two  external parameters, namely, the concentration  of ATP in the environment (that regulates the chemical potential difference $\Delta \mu$ and is denoted by denoted by ${[ATP]}$),  and the mechanical force $f$ experienced by the motor.  
The transitions $3\rightarrow4$ and $1\rightarrow2$ represent the binding of ATP to one of the motor heads ,and therefore 
\begin{equation}
\mathbf{q}_{34} = \mathbf{q}_{12} = \mathbf{q}^0_{12}  {[ATP]},
\end{equation}
where $\mathbf{q}^0_{12} $ is a constant.  
The transitions $4\rightleftharpoons2$,$2\rightleftharpoons3$ and $3\rightleftharpoons1$ all correspond to the detachment or attachment of one of the heads of the molecular motor from the microtubule, which is the ``stepping" of the motor. These transitions are thus affected by the mechanical force $f$ which modifies their rates according to 
\begin{equation}
\mathbf{q}_{xy} = \mathbf{q}^0_{xy} \exp\left(-\delta_{xy} \frac{fd}{k_{\rm_B}\mathsf{T}_{\rm env}}\right)
\end{equation}
for transitions corresponding to forward steps and 
\begin{equation}
\mathbf{q}_{xy} = \mathbf{q}^0_{xy} \exp\left(\delta_{xy} \frac{fd}{k_{\rm_B}\mathsf{T}_{\rm env}}\right)
\end{equation} 
for transitions corresponding to backward steps.  Here the  force distribution factors $\delta_{xy}$ are non-negative real parameters that determine the extent to which the transition from $x$ to $y$ is affected by the mechanical force $f$.   In our simulations we use the values of the load distribution factors and the transition rates from Ref.~\cite{shen2022mechanochemical} that obtained these parameters from fits to  to experimental data; we specify all parameters used in   Appendix~\ref{sec:fig3params}.

We now fix the coefficients $c_{xy}$ defining the positional current $J_{\rm pos}$.  Each step of the motor, of length $d$, consists of two substeps both forward stepping cycles. Since the total step size in both cycles is the same, we get the equality
\begin{equation}
    c_{23} + c_{21} =  c_{23} + c_{42} = 1,
\end{equation}
where we have normalized the coefficients such that the current $J_{\rm pos}$ is non-dimensionalized.  Next, we use thermodynamic consistency to relate the coefficients $c_{xy}$ to the force distribution factors $\delta_{xy}$.    On one hand, the dissipation in the environment due to the work by the motor on the external load  when the motor moves from state $x$ to state $y$ equals  
\begin{equation}
  \frac{f  }{k_{\rm B}\mathsf{T}_{\rm env}}  c_{xy} d.  \label{eq:75}
\end{equation}
On the other hand, due to local detailed balance the entropy change in the environment is given by 
\begin{equation}
\ln \frac{\mathbf{q}_{xy}}{\mathbf{q}_{yx}} = \ln \frac{\mathbf{q}^{(0)}_{xy}}{\mathbf{q}^{(0)}_{yx}} + \frac{fd}{k_{\rm B}\mathsf{T}_{\rm env}} (\delta_{xy}+\delta_{yx}).\label{eq:76}
\end{equation}
Since the second term in Eq.~(\ref{eq:76}) quantifies the dissipation due to the external force $f$, thermodynamic consistency implies that it equals to (\ref{eq:75}), yielding the  thermodynamic consistency condition  
\begin{equation}
    c_{xy} = \delta_{xy} + \delta_{yx}\label{eq:thermocons}
\end{equation} 
that relates the current coefficients to the force distribution factors. 

\subsection{Relationship between effective affinity and the stalling force}\label{sec:EASta}
\begin{figure}
\centering
\tabskip=0pt
\includegraphics[width=0.8\textwidth]{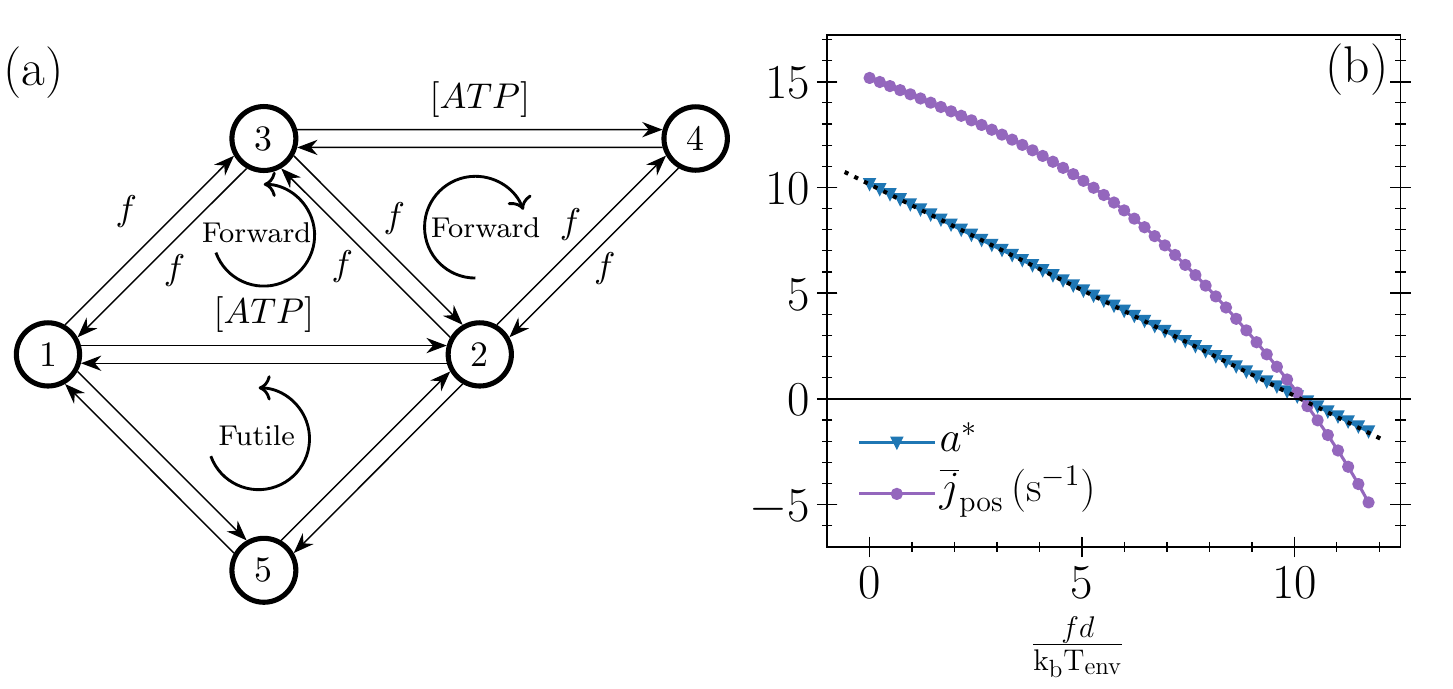}
    \caption{Panel (a): Graph of admissible transitions in the  mechanochemical model    of  Kinesin-1 from Ref.~\cite{shen2022mechanochemical}.  The model has three  cycles, two  corresponding with forward motion and one futile cycle.   The dependence of the rates on the mechanical forces $f$ and $[ATP]$ are as indicated.       (b) The average positional current $\overline{j}$ (purple circles) and effective affinity $a^\ast$ (blue triangles) as a function of $f$.   The effective affinity $a^\ast$ is obtained by numerically diagonalising $\mathbf{\tilde{q}}(a)$ and numerically finding the value of $a$ for which the largest eigenvalue equals zero.   The markers for   $\overline{j}$ are obtained by numerically calculating the steady state of the Markov process through a linear solver. See  Appendix~\ref{sec:fig3params} for the model parameter values. The black dashed line on top of the blue triangles shows  Eq.~\eqref{eq:toverify} for this model.} 
    \label{fig:4}
\end{figure}

\begin{figure}[t!]
 \centering
 \includegraphics[width=0.85\textwidth]{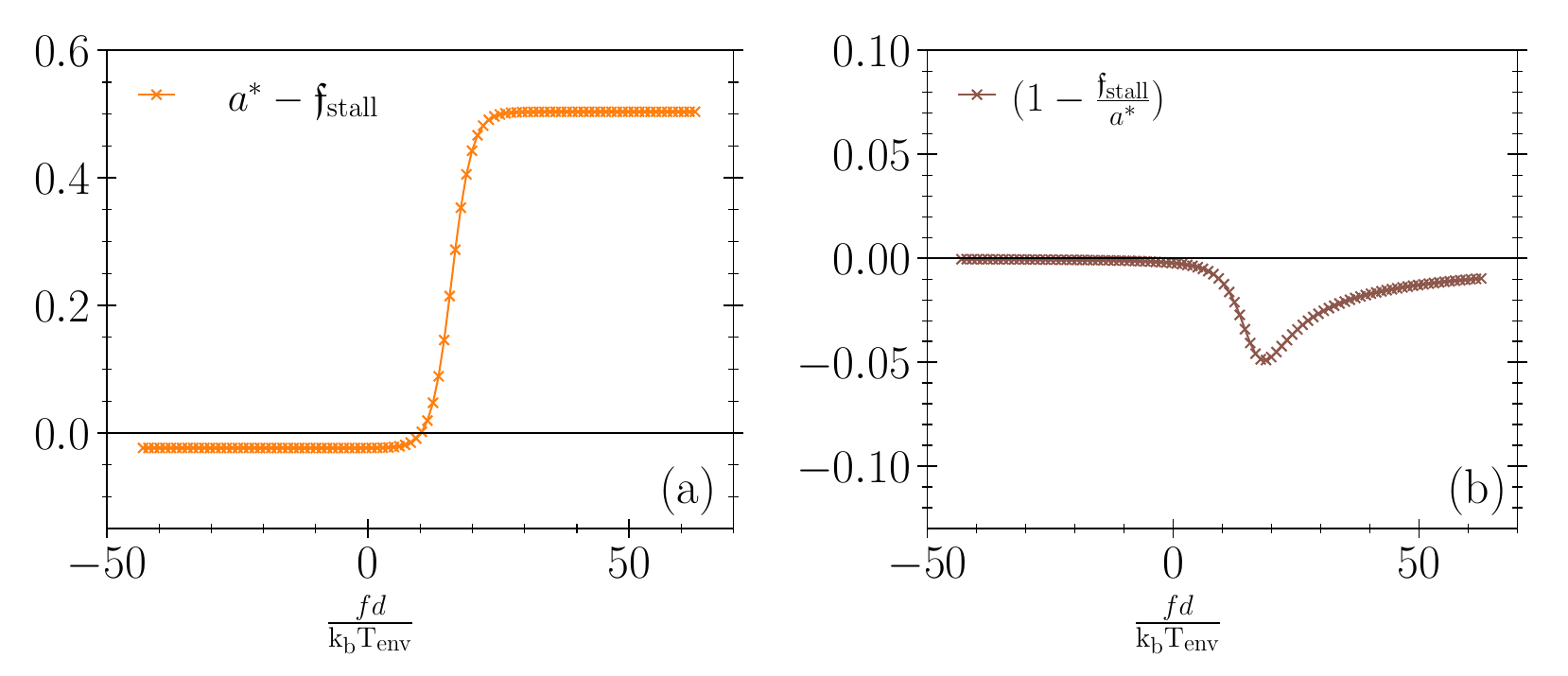}
 \caption{{\it Difference between the effective affinity $a^\ast$ and the stalling force $\mathfrak{f}_{\rm stall}$} for the model of Kinesin-1 depicted in Fig.~\ref{fig:4}(a). --- The left panel (a) shows the difference between the left and right hand sides of Eq~\eqref{eq:toverify}. The right panel (b) shows same quantity, now divided by $a^\ast$ to show the relative difference. The model parameters used are the same as for Fig.\ref{fig:4}(b), and are mentioned in Appendix.~\ref{sec:fig3params}}
\label{fig:5}   
\end{figure}

We now address the question of whether the effective affinity in the present model is a stalling force, i.e., whether Eq.~(\ref{eq:toverify}) applies in this model. In Fig~\ref{fig:4}(b), we plot $a^\ast$ as a function of $f$ and compare it with a plot of Eq~\eqref{eq:toverify}.   Remarkably, the two appear to correspond well.  However, upon closer examination we find a systematic deviation from the linear behaviour of Eq.~\eqref{eq:toverify}.  The systematic deviation is clear from  Fig.~\ref{fig:5}(a) that shows the difference between the effective affinity $a^\ast$ and the dimensionless form of the stalling force $\mathfrak{f}_{\rm stall}$ as a function of $f$.   The difference $a^\ast-\mathfrak{f}_{\rm stall}$ is a bounded function that saturates for large values of $f$, and is small compared to $a^\ast$ (the ratio $1-\mathfrak{f}_{\rm stall}/a^\ast\approx 10^{-2}$, see Fig.~\ref{fig:5}(b)).   Hence, although the effective affinity is not equal to the stalling force,  we nevertheless find that in this model the stalling force closely approximates the effective affinity.

The  numerical observation that the effective affinity is approximately equal to the stalling force, is closely related to the fact that the model is thermodynamically consistent, in the sense that Eq.~(\ref{eq:thermocons}) holds.    Indeed, if we modify the model such that Eq.~(\ref{eq:thermocons}) is not anymore valid, then the effective affinity deviates  significantly   from the stalling force (see Appendix~\ref{sec:nothermocons}).   In addition, the approximate linear scaling of $a^\ast$ with $f$ with slope $-1$ is clearly violated for models for which the thermodynamic consistency condition (\ref{eq:thermocons})  does not hold (results not shown).

\section{Discussion}\label{sec:discussion}
Complex nonequilibrium systems, such as the metabolic pathways in living cells, are often governed by a large number of   currents that are coupled with each other. The coupling makes it difficult to predict the properties of individual currents (such as their direction), as the external constraints imposed by the thermodynamic affinities are insufficient to determine the currents' properties.   In this paper, we have introduced  an affinity-like quantity, called the effective affinity, that extends the properties of thermodynamic affinities to coupled currents in   complex nonequilibrium systems.  

The effective affinity is a (unique) real number associated with a fluctuating current that quantifies  several properties of currents that physicists and chemists usually assign to thermodynamic affinities of uncoupled currents. Notably, the effective affinity determines the sign of the associated current, and the effective affinity multiplied by the average current lower bounds the rate of dissipation as expressed by the inequality (\ref{eq:lowerbound}).  Furthermore,   the effective affinity determines the fluctuations of a current, as it is  the exponential decay constant that characterises the tails of the infimum statistics of the current [see  Eq.~(\ref{eq:inf})].  If all fundamental currents of a Markov process are uncoupled, then the effective affinity of a current is equal to the current's thermodynamic affinity that appears in the decomposition (\ref{eq:meso}) of entropy production.       In addition, we have shown that all fluctuating currents that are cycle equivalent with $kS_t$, with $k$ a constant, are optimal in the sense that for those currents the equality in (\ref{eq:lowerbound}) is attained. 

From a methodological point of view, this paper introduces  a new class of  martingales $M_t$, given by Eq.~(\ref{eq:Mt}), that are associated with generic currents.    The martingale $M_t$ is an exponential martingale, and the prefactor that appears in the exponential in front of the current is the negative effective affinity.    Martingales considered before in the literature, such as the  exponentiated negative entropy production~\cite{chetrite2011two, neri2017statistics, neri2019integral} and the exponential martingale for  edge currents derived in Ref.~\cite{neri2023extreme}, are special cases of the martingale $M_t$.        Given the numerous  properties of martingales, as outlined in~\cite{roldan2022martingales}, the exponential martingale $M_t$ adds to   existing  techniques for studying current fluctuations.  Notably, in this Paper we  have used $M_t$ to derive the thermodynamic trade off relation Eq.~(\ref{eq:ineq2})  that expresses an inequality between speed (quantified by $\langle T\rangle$), uncertainty (quantified by $p_-$),and dissipation (quantified by $\dot{s}$).   The derivation in this paper based on martingales is different from the one in~Ref.~\cite{neri2022universal}, and the present derivation is arguably more clear as it does not  rely on a scaling argument.   Another interesting aspect of the martingales $M_t$ is that they relate large deviation theory (as $a^\ast$ is the nonzero root of  the equation $\lambda_J(a)=0$) with the  extreme value statistics of the current (as $a^\ast$ is the exponential decay constant in the tails of the infimum statistics of $J$).   

For edge currents, the effective affinity $a^\ast$  equals   the  effective affinity  studied in Refs.~\cite{polletini1,polettini2019effective} for this case.    We have shown that several properties of the edge effective affinity generalise to the effective affinity defined in this paper.   However,  there are a few  properties  that we have not explored in this Paper.    Notably, Ref.~\cite{pedro1, garilli2023fluctuation} derives a  detailed fluctuation relation for currents,  which holds as long as the distribution is evaluated at the random time when a fixed number of transitions have been observed, and Ref.~\cite{harunari_what_2022}   shows that the effective affinity determines the Kullback-Leibler divergence due to subsequent transitions along an edge.    How these properties  extend to the case of  generic currents   remains  an open problem.

The last three sections (Secs.~\ref{sec:twocycle}, \ref{sec:diffusion}, and \ref{sec:kinesin}) are case studies for   three important problems.   We discuss  here how these problems can be generalised. 

An interesting open problem is to identify the fluctuating currents that are optimal, in the sense that for these currents   the equality in  (\ref{eq:ineq2}) is attained.  We know that the set of optimal currents contains the sets $[kS_t]$ of currents that are cycle equivalent with the stochastic entropy production $S_t$, up to a proportionality constant $k$.     However, it remains open whether there exist currents that are optimal even though they are not in one of the sets $[kS_t]$.    Numerical analysis in simple models, such as the  Markov jump process in Sec.~\ref{sec:twocycle}, indicate that all optimal currents are contained in the sets $[kS_t]$.   However, we do  not have a  mathematical argument to support this numerical observation, and therefore this interesting question remains open.  

Although we have focused in this paper on time-homogeneous Markov processes defined on a discrete set, the definition of the effective affinity as the nonzero root of $\lambda_J(a)=0$ also applies to overdamped Langevin processes.   To verify this claim, we have analysed in Sec.~\ref{sec:diffusion} the effective affinity in a Brownian particle bound on a ring that is subject to a constant external force. We have shown  that for any current in this model the effective affinity is the thermodynamic force acting on the current, and therefore all currents are optimal.  We have also derived an explicit expression for the martingale $M_t$ of a generic fluctuating current.    Extending the present theory to generic overdamped Langevin processes requires us to extend the formalism of the present paper to this setup, and will be considered in future work.  It remains to be determined how generally applicable effective affinities are in the context of overdamped Langevin processes.   In addition, it will be interesting to extend the definition of  cycle equivalence classes to  Langevin processes; the  framework of  Refs.~\cite{qian1998vector, yang2021bivectorial} that defines   cycle currents in Langevin processes as 2-forms may be useful to this purpose.             

This brings us to the last case study, which investigates the relationship between the effective affinity and the stalling force.  Since the effective affinity equals zero when the current is zero, it can in general not be identified with the thermodynamic force (except for unicyclic systems, such as a Brownian particle bound on a ring), but there may still be a connection with the stalling force.          In Sec.~\ref{sec:kinesin}, we have numerically analysed the effective affinity and stalling force of a biophysical model of Kinesin-1.   If this model is thermodynamically consistent, then the effective affinity and the stalling force are approximatly equal (the relative difference $1-\mathfrak{f}_{\rm stall}/a^\ast\approx 10^{-2}$).  On the other hand, if the thermodynamic consistency condition (\ref{eq:thermocons}) is violated, then we have observed a significant difference between the effective affinity $a^\ast$ and the stalling force $\mathfrak{f}_{\rm stall}$.    Hence, for this biophysical model we can conclude that for practical purposes the stalling force can be used to estimate the effective affinity.   This observation is interesting as it relates  via Eq.~(\ref{eq:inf}) the exponential decay constant of the infima of the positional current of   Kinesin-1 motors  to the stalling force.   In addition,  the stalling force can be used to estimate the rate of dissipation through 
$\mathfrak{f}_{\rm stall}\overline{j}$.  
It remains to be understood whether the close relationship between the effective affinity and the stalling force is a specific property of this model, or whether it is generally applicable to thermodynamically consistent models.   

\section*{Acknowledgments} 
We thank  Nikolas N\"{u}sken for discussions and guidance,  we  thank Stefano Bo and  Simone Pigolotti for a detailed reading of the manuscript, and we thank Lennart Dabelow, Rosemary Harris,  Friedrich H\"{u}bner,  Anupam Kundu,  Christian Maes, Gabriele Pinna, and Alvaro Lanza Serrano  for fruitful discussions.

\begin{appendices}
\numberwithin{equation}{section}
\section{Model parameters for Figure 1}\label{sec:fig1params}
Figure 1 shows $\lambda_J$ and $\lambda_{S'}$ for the Markov jump model with a graph of admissible transitions as illustrated  in Figure 2.     The transition rates between the states are set equal to   $\mathbf{q}_{12}=3$, 
$\mathbf{q}_{41}=3$, 
$\mathbf{q}_{21}=1$, 
$\mathbf{q}_{14}=1$, 
$\mathbf{q}_{23}=2$, 
$\mathbf{q}_{34}=2$, 
$\mathbf{q}_{32}=1$, 
$\mathbf{q}_{43}=1$, 
$\mathbf{q}_{24}=3$, and 
$\mathbf{q}_{42}=1$, and the current $J$   is defined by the coefficients $c_{12} = -c_{21} = 1.26913$, $c_{34} = -c_{43} = 1.37403$, and  all other $c_{x,y}$ coefficients are set to $0$. 
For the abovementioned rates the  entropy production rate $\dot{s} = 1.94745$; notice that in the figure this is equal to the  nonzero root of $\lambda_{S'}$.  
  The curves for $\lambda_J$ and $\lambda_{S'}$ are obtained by numerically diagonalising the corresponding matrix $\tilde{\mathbf{q}}$.    The current $S'$ is the current with  coefficients   $c_{xy} = \frac{1}{\dot{s}} {\rm ln} \left(\frac{\mathbf{q}_{xy}}{\mathbf{q}_{yx}}\right)$.

\section{Effective affinity for edge currents}
Here we consider the limiting case of   edge currents $J_t = J^{xy}_t$ that count the net number of jumps along an edge $(x,y)$ of the graph of admissible transitions of the Markov chain $X$.    In Sec.~\ref{sec:S6} we show that the effective affinity for edge currents takes the form given in Eq.~(\ref{eq:effectEdge}), and in Sec.~\ref{subsec:edgecurrentmartingale} we show that the Martingale $M_t$, as defined in Eq.~(\ref{eq:Mt}),  is equivalent to the martingale in Eq.~(69) of Ref.~\cite{neri2023extreme}.
\subsection{Effective affinity for edge currents}\label{sec:S6}
We show that for edge currents the effective affinity is given by 
\begin{equation}
a^\ast = \ln \frac{p^{(x,y)}_{\rm ss}(x)\mathbf{q}_{xy}}{p^{(x,y)}_{\rm ss}(y) \mathbf{q}_{yx}},
\label{eq:edgeea}
\end{equation}
where the right-hand side of (\ref{eq:edgeea}) is the edge current effective affinity defined in Refs.~\cite{polletini1,polettini2019effective, neri2023extreme}, and the left-hand side is the effective affinity as defined in Eq.~(\ref{eq:defa}). 

In Eq.~(\ref{eq:edgeea}), the $p^{(x,y)}_{\rm ss}$ is the stationary distribution of a Markov  jump process with the $\mathbf{q}$-matrix  $\mathbf{q}^{(x,y)}$ that has  off-diagonal entries  given by 
\begin{equation}
    \mathbf{q}^{(x,y)}_{rs} =  
    \begin{cases}
        0 ,\quad &  (r,s) \in \{(x,y) , (y,x)\}, \\
        \mathbf{q}_{rs}, \quad & (r,s) \in \mathcal{X}^2 \setminus  \left(\{(x,y) , (y,x)\} \cup \left\{(z,z):z\in \mathcal{X}\right\}\right),\\
    \end{cases}
    \label{eq:qdel}
\end{equation}
and  diagonal entries given by  
\begin{equation}
 \mathbf{q}^{(x,y)}_{rr} = -\sum_{z \in \mathcal{X} \setminus \left\{r\right\}}   \mathbf{q}^{(x,y)}_{rz}.  
\end{equation}
We  assume this Markov process is  ergodic, which implies that $p^{(x,y)}_{\rm ss}$ is a strictly positive eigenvector of $\mathbf{q}^{(x,y)}$ with eigenvalue zero.

By definition, the effective affinity $a^\ast$  is the value of $a$ at which  the Perron root of  $\tilde{\mathbf{q}}(a)$ equals zero.  
In the present example of an edge current, the  tilted matrix  $\tilde{\mathbf{q}}(a)$ is given by Eq.~(\ref{eq:qtilde})  with $c_{xy} = - c_{yx} = 1$ and all other $c$-coefficients equal to $0$, leading to the expression  
\begin{equation}
    \tilde{\mathbf{q}}_{rs}(a) =  
    \begin{cases}
        \mathbf{q}_{xy} e^{-a} ,\quad &   (r,s) = (x,y),\\
        \mathbf{q}_{yx} e^{a} ,\quad &  (r,s) = (y,x),\\
        \mathbf{q}_{rs}, \quad & (r,s)  \in \mathcal{X}^2 \setminus  \left(\{(x,y) , (y,x)\} \cup \left\{(z,z):z\in \mathcal{X}\right\}\right),\\
        -\sum_{z \in \mathcal{X} \setminus \left\{r\right\}}   \mathbf{q}_{rz}, \quad & r=s.\\
    \end{cases}
    \label{eq:qtildeedge}
\end{equation}

To derive  Eq.~\eqref{eq:edgeea}, we need to show that if $a$ takes the value in the right-hand side of Eq.~\eqref{eq:edgeea}, then the Perron root of the matrix $\tilde{\mathbf{q}}(a)$ vanishes.   This latter follows from the fact that: (i) when $a$ equals  the right-hand side of Eq.~\eqref{eq:edgeea}, then  $p^{(x,y)}_{\rm ss}$  is  a left eigenvector of $\tilde{\mathbf{q}}(a)$ with eigenvalue zero; (ii) any strictly positive left eigenvector of a positive matrix is a left eigenvector of its Perron root,  see Theorem 1.4 of Ref.~\cite{berman1994nonnegative}.

In what follows, we show that   if $a$ equals the right-hand side of Eq.~(\ref{eq:edgeea}), then $p^{(x,y)}_{\rm ss}$ is a left eigenvector of  $\tilde{\mathbf{q}}(a)$ with zero eigenvalue, i.e., we demonstrate that 
\begin{equation}
    \sum_{s\in \mathcal{X}\setminus \left\{r\right\}}\tilde{\mathbf{q}}_{sr}(a)p^{(x,y)}_{\rm ss}(s)  - \tilde{\mathbf{q}}_{rr}(a)p^{(x,y)}_{\rm ss}(r) = 0, \qquad \forall r \in \mathcal{X}  \label{eq:S5}
\end{equation}
holds for the corresponding value of $a$.  

For  $r \in \mathcal{X}\setminus \{x,y\}$   Eq.~(\ref{eq:S5}) is identical to 
\begin{equation}
\sum_{s\in \mathcal{X}\setminus \left\{x\right\}}\mathbf{q}_{sr}p^{(x,y)}_{\rm ss}(s)  - \mathbf{q}_{rr}p^{(x,y)}_{\rm ss}(r) = 0,  \label{eq:S6x}
\end{equation}
which we recognise as the stationary condition that defines $p^{(x,y)}_{\rm ss}$, and is thus satisfied.   For $r=x$,   Eq.~(\ref{eq:S5}) is identical to 
\begin{eqnarray}
     &&    \sum_{s\in \mathcal{X}\setminus \left\{x\right\}}\tilde{\mathbf{q}}_{sx}(a)p^{(x,y)}_{\rm ss}(s)  - \tilde{\mathbf{q}}_{xx}(a)p^{(x,y)}_{\rm ss}(x) = 0 \label{eq:S6}\\
  \Leftrightarrow &&       \mathbf{q}_{yx} e^{a} p^{(x,y)}_{\rm ss}(y) - {\mathbf{q}}_{xy}p^{(x,y)}_{\rm ss}(x) = 0 \label{eq:S7} \\
    \Leftrightarrow &&       a = \ln \frac{p^{(x,y)}_{\rm ss}(x)\mathbf{q}_{xy}}{p^{(x,y)}_{\rm ss}(y) \mathbf{q}_{yx}},
\end{eqnarray}
which yields the assumed value of $a$.   Equation~(\ref{eq:S7}) follows from (\ref{eq:S6}), (\ref{eq:qtildeedge}), and (\ref{eq:S6x}) for $r=x$.   Analogously one can show that Eq.~(\ref{eq:S5}) holds for $r=y$.

\subsection{Martingale for edge currents}\label{subsec:edgecurrentmartingale}
We  show that for edge currents the martingale $M_t$ from Eq.~\eqref{eq:Mt}  is equivalent with the martingale 
\begin{equation}
M^{xy}_t := \frac{p_{\mathrm{ss}}^{(x, y)}(X_0) r_{\mathrm{ss}}(X_t)}{p_{\mathrm{ss}}^{(x, y)}(X_t) p_{\mathrm{ss}}(X_0)} \exp\left(-a^* J^{x  y}_t\right),\label{eq:edgemartingale1}
\end{equation}
defined in Ref.~\cite{neri2023extreme},  in the sense that the two martingales are equal up to a  (random)  prefactor that is constant in time.      In Eq.~(\ref{eq:edgemartingale1})  $r_{\rm ss}$ is the steady state of the  modified Markov process with rate matrix  
\begin{equation}
\mathbf{l}_{u   v}= \begin{cases}\mathbf{q}_{u   v}, & \text { for }(u, v) \in\{(x, y),(y, x)\}, \\ \frac{p_{\rm ss}^{(x,y)}(v)}{p^{(x,y)}_{\rm ss}(u)} \mathbf{q}_{v   u}, & \text { for }(u, v) \in \mathcal{X}^2 \backslash\{(x, y),(y, x)\}.\end{cases}
\end{equation}  

Indeed, as we show in this appendix, 
\begin{equation}
M^{xy}_t = \frac{p_{\mathrm{ss}}^{(x, y)}(X_0)}{ p_{\mathrm{ss}}(X_0)} M_t. \label{eq:equiv}
\end{equation}  
The equivalence expressed by  (\ref{eq:equiv}) follows from using the identity 
\begin{equation}
\phi_{a^{\ast}}(z) = \frac{r_{\rm ss}(z)}{p^{(x,y)}_{\rm ss}(z)},  \label{eq:phiastedge}
\end{equation}
which holds for  all $z\in \mathcal{X}$, in Eq.~\eqref{eq:Mt} and comparing the resultant expression with \eqref{eq:edgemartingale1}.   

In what follows we derive Eq.~(\ref{eq:phiastedge}).   Consider the generator $\mathbf{m}$ defined by 
\begin{equation}
    \mathbf{m}_{uv}= \begin{cases}\frac{r_{\rm{ss}}(v)}{r_{\rm{ss}}(u)}\mathbf{q}_{vu}, & \text { for }(u, v) \in\{(x, y),(y, x)\}, \\ \frac{r_{\rm{ss}}(v)}{r_{\rm{ss}}(u)}\frac{p_{\rm ss}^{(x,y)}(v)}{p^{(x,y)}_{\rm ss}(u)} \mathbf{q}_{uv}, & \text { for }(u, v) \in \mathcal{X}^2 \backslash\{(x, y),(y, x)\},\end{cases}\label{eq:timereverersel}
\end{equation}
which describes the time-reversal  of the process described by $\mathbf{l}$.
Note that the processes governed by $\mathbf{l}$ and $\mathbf{m}$ have the same steady-state distribution  $r_{\rm ss}$ due to the invariance of the steady state distribution under time reversal.

Using the expression (\ref{eq:edgeea}) for the effective affinity of an edge current, we can express the matrix $\mathbf{m}$ as 
\begin{equation}
\mathbf{m} = \mathbf{r_{\rm ss}}^{-1}\mathbf{p}^{(x,y)}_{\rm ss} \tilde{\mathbf{q}}(a^\ast)\mathbf{p^{(x,y)}_{\rm ss}}^{-1}\mathbf{r_{\rm ss}},   \label{eq:l}
\end{equation} 
where $\mathbf{r_{\rm ss}}$ and $\mathbf{p}^{(x,y)}_{\mathrm ss}$ are diagonal matrices with entries given by $r_{\rm ss}(z)$ and ${p}^{(x,y)}_{\rm ss}(z)$, respectively, and where $\mathbf{r_{\rm ss}}^{-1}$ and $\mathbf{p^{(x,y)}_{\rm ss}}^{-1}$ are matrix inverses.   The  matrix $ \tilde{\mathbf{q}}(a^\ast)$ is given by (\ref{eq:qtilde})  with $c_{xy}=-c_{xy}=1$, and all other coefficients $c$ are set to zero.   

We compare $\mathbf{m}$ with the matrix 
\begin{equation}
\mathbf{n} = \boldsymbol{\phi^{-1}_{a^\ast}} \tilde{\mathbf{q}}(a^\ast)\boldsymbol{\phi_{a^\ast}},   \label{eq:l2}
\end{equation}
where  $\boldsymbol{\phi}_{a}$ is the diagonal matrix with entries given by  the  entries $\phi_{a}(x)$  of the right eigenvector  associated with the Perron root of $\tilde{\mathbf{q}}(a)$.     The matrix $\mathbf{n}$ is a special case of the  generalized Doob transform~\cite{touchette2009large} 
\begin{equation}
\boldsymbol{\phi^{-1}_{a}} \tilde{\mathbf{q}}(a)\boldsymbol{\phi_{a}}  - \lambda(a)\mathbbm{1}_{|\mathcal{X}|} ,
\end{equation}
where $\mathbbm{1}_{|\mathcal{X}|}$ is the identity matrix of order $|\mathcal{X}|$.

Both $\mathbf{m}$ and $\mathbf{n}$ are diagonal transformations of the tilted matrix $\tilde{\mathbf{q}}(a^\ast)$.  In addition, both matrices  $\mathbf{m}$ and  $\mathbf{n}$ are Markov matrices. It follows then the diagonal matrices in the corresponding transformations are proportional. Setting the proportionality constant to $1$, we obtain Eq.~\eqref{eq:phiastedge}. 

We demonstrate the aforementioned proportionality in what follows. Since $\mathbf{m}$ and $\mathbf{n}$ are Markov matrices, we have
\begin{equation}
\mathbf{m} \cdot \vec{1} = \mathbf{n} \cdot \vec{1} = 0.  
\end{equation}
where $\vec{1}$ is the column vector with all entries equal to $1$. Then from Eqs.~(\ref{eq:l}) and (\ref{eq:l2}) it follows that
\begin{equation}
    \boldsymbol{\phi^{-1}_{a^\ast}} \tilde{\mathbf{q}}(a^\ast) \cdot \phi_{a^\ast} = \mathbf{r_{\rm ss}}^{-1}\mathbf{p}^{(x,y)}_{\rm ss} \tilde{\mathbf{q}}(a^\ast) \cdot ({p^{(x,y)}_{\rm ss}}^{-1} \circ r_{\rm ss}) = 0 ,
    \label{eq:eigenvec}
\end{equation}
where $\phi_{a^\ast}$, ${p^{(x,y)}_{\rm ss}}^{-1}$ and $r_{\rm ss}$ are column vetors and $\circ$ is the element-wise product of the two column vectors. Since $\boldsymbol{\phi^{-1}_{a^\ast}}$,$\mathbf{r_{\rm ss}}^{-1}$ and $\mathbf{p}^{(x,y)}_{\rm ss}$ are diagonal matrices with positive diagonal entries, Eq. \eqref{eq:eigenvec} implies that $\phi_{a^\ast}$ and  $({p^{(x,y)}_{\rm ss}}^{-1} \circ r_{\rm ss})$ are right eigenvectors of $\tilde{\mathbf{q}}(a^\ast)$ with eigenvalue zero (in the case of $\phi_{a^\ast}$, this is also true by definition). Thus the two vectors must be equal up to a proportionality constant which we set to $1$, giving Eq. \eqref{eq:phiastedge} which we were meant to derive.

\section{Wald's equation for fluctuating currents}\label{app:Wald}

We derive the Eq.~(\ref{eq:Wald}) relating the mean first passage time $\langle T\rangle$ to the current $\overline{j}$, and we clarify the relation between  (\ref{eq:Wald}) and  Wald's equation~\cite{Wald1,Wald2}.   

Let us denote the scaled cumulant generating function  of $T$ conditioned on escape through the positive threshold by 
\begin{equation}
m_+(\mu) := \lim_{\ell_{\rm min}\rightarrow \infty} \frac{\ln \langle e^{\mu T} \rangle_+}{\ell_+},
\end{equation}
where $\langle \cdot \rangle_+$ is an expectation conditioned on  the event $J_T\geq \ell_+$.  As shown in Ref.~\cite{gingrich2017fundamenta}, $m_+$ is the functional inverse of $\lambda_J$, i.e.,
\begin{equation}
\lambda_J(m_+(\mu)) = -\mu.\label{eq:mP}
\end{equation}  
Using  
\begin{equation}
\lambda_J(a) = -\overline{j}a + O(a^2)
\end{equation}
and  
\begin{equation}
m_+(\mu) = \mu \lim_{\ell_+\rightarrow \infty}\frac{\langle T\rangle_+}{\ell_+} + O(\mu^2)
\end{equation}
in Eq.~(\ref{eq:mP}), and using that  (\ref{eq:mP}) holds for all values of $\mu$ in the neighbourhood of the origin, we find that 
\begin{equation}
\overline{j} \lim_{\ell_+\rightarrow \infty} \frac{\langle T\rangle_+}{\ell_+}  = 1. \label{eq:resultWald}
\end{equation}
Considering that $p_+\rightarrow 1$ as $\ell_{\rm min}\rightarrow \infty$, we can  set $\langle T\rangle = \langle T\rangle_+ (1+o_{\ell_{\rm min}}(1))$ in  
(\ref{eq:resultWald}), yielding   the Eq.~(\ref{eq:Wald})  that we were meant to derive.  

We call Eq.~(\ref{eq:Wald}) Wald's equation for fluctuating currents~\cite{Wald1,Wald2}, as it can be seen as an asymptotic version of Wald's equation.  Indeed,   setting $\ell_+ = \langle J_T\rangle(1+o_{\ell_{\rm min}}(1))$ in  Eq.~(\ref{eq:Wald}) gives
\begin{equation}
\langle J_T\rangle = \langle T\rangle \overline{j} (1+o_{\ell_{\rm min}}(1)),
\end{equation}
which is reminiscent of Wald's equation for sums of independent and identically distributed random variables~\cite{Wald1,Wald2}.

\section{Currents in the same cycle equivalence classes have the same  effective affinity}\label{sec:S5}
     
In this appendix we derive  Eq.~\eqref{eq:charpol1} that expresses the characteristic polynomial of the matrix  $\mathbf{\tilde{q}}(a)$ as a sum over spanning, linear subgraphs of the graph of admissible transitions of the Markov chain $X$.  

The appendix is structured as follows.  First, in Sec.~\ref{sebsec:fundamentalbasis} we  define  fundamental cycle basis $\mathcal{C}$ of a graph, which we  use   in Sec.~\ref{sec:recap} to define the    cycle coefficients $c_\gamma$ and the cycle equivalence classes $[J_t]$.   In the following three sections, we derive the expression  Eq.~\eqref{eq:charpol1}. First, in  Sec.~\ref{subsec:S5A}, we use the Coefficients  Theorem for directed graphs to express the characteristic polynomial of $\tilde{\mathbf{q}}(a)$ as  a sum over linear  subgraphs.    Second, in Sec.~\ref{subsec:S5BB} we define the coefficients $c_{\mathscr{C}}$ associated with directed cycles  $\mathscr{C}$, and we show that these coefficients can be expressed as a linear combination of the  cycle coefficients $c_\gamma$.    Lastly, in Sec.~\ref{subsec:S5B} we combine the above arguments to obtain  Eq.~\eqref{eq:charpol1}.   

\subsection{Fundamental cycle basis}\label{sebsec:fundamentalbasis}

We construct a set of cycles $\mathcal{C}$, known as a fundamental cycle basis,  associated with a nondirected graph $\mathcal{G} = (\mathcal{X},\mathcal{E})$, where $\mathcal{X}$ is the set of vertices and $\mathcal{E}$ the set of nondirected edges.  
A cycle is an ordered  sequence  $\gamma = [x_1,x_2,\ldots, x_{n(\gamma)},x_1]$ of nodes $x_i \in \mathcal{X}$ such that $(x_{i},x_{i+1})\in \mathcal{E}$ for all $i$ and each vertex is only traversed once; notice that reversing the sequence leaves the cycle invariant.    A cycle basis is a minimal set of cycles, called basis cycles, such that any cycle $\gamma$ of $\mathcal{G}$ can be expressed as a symmetric difference of basis cycles. The set $\mathcal{C}$, known as a fundamental cycle basis of the graph $\mathcal{G}$ is a cycle basis that can be constructed from a spanning tree $\mathcal{T}$ of $\mathcal{G}$.
The spanning tree $\mathcal{T}$ has a number $|\mathcal{X}|-1$ of edges, and hence there are a number $|\mathcal{E}|-|\mathcal{X}|+1$  of edges in  $\mathcal{E}$ that do not belong to the spanning tree.   Adding one new edge to a spanning tree creates a graph with one cycle, and in this way we can define a number   $|\mathcal{E}|-|\mathcal{X}|+1$  of  cycles $\gamma \in \mathcal{C}$, and the set $\mathcal{C}$ forms a basis of the cycle space, which we refer to as a fundamental cycle basis. See also~Ref.~\cite{baratoSymmetryCurrentProbability2012, bollobas_modern_1998} for more detailed definitions.

In Sec.~\ref{sec:recap} we use a fundamental cycle basis $\mathcal{C}$ of the graph of admissible transitions of a Markov Jump process to define the coefficients $c_\gamma$ associated with the basis cycles $\gamma \in \mathcal{C}$ as 
\begin{equation}
c_\gamma:=\sum_{i=1}^{n(\gamma)} c_{x_i x_{i+1}}, \quad \gamma \in \mathcal{C},
\end{equation}
where $n(\gamma)$ is the number of nodes in $\gamma$. We arbitrarily fix a direction in which to perform this sum for each $\gamma$, since reversing the order of nodes in a basis cycle $\gamma$ does not alter the cycle, but would change the sign of $c_\gamma$ as $c_{xy} = -c_{yx}$.

\subsection{Coefficients Theorem applied to $\tilde{\mathbf{q}}(a)$}\label{subsec:S5A}
We now apply the Coefficients Theorem for directed graphs, see Refs.~\cite{acharya_spectral_1980,cvetkovic_spectra_1995},  to the matrix   $(\tilde{\mathbf{q}}(a) - \xi \mathbbm{1})$, where $\mathbbm{1}$ is the identity matrix of size $\mathcal{X}$.  The Coefficients Theorem gives an expression for the determinant of the adjacency  matrix of a weighted graph in terms of its  spanning linear subgraphs.    To apply this theorem, we interpret the matrix $(\tilde{\mathbf{q}}(a) - \xi \mathbbm{1})$ as the adjacency matrix of a weighted directed graph for which  $\mathcal{X}$ is the set of nodes and  the nonzero entries of   $(\tilde{\mathbf{q}}(a) - \xi \mathbbm{1})$ determine the weights of the directed edges in the graph.   Thus, we obtain a graph as illustrated for an example in Panel (a) of Fig.~\ref{fig:2}. 

Using the Coefficients Theorem for a weighted graphs, see Refs.~\cite{acharya_spectral_1980,cvetkovic_spectra_1995} for a detailed description, we  express the characteristic polynomial of $\tilde{\mathbf{q}}(a)$ as 
\begin{equation}
    {\rm det} \left(\tilde{\mathbf{q}}(a) - \xi \mathbbm{1} \right) =  (-1)^{\vert\mathcal{X}\vert}\sum_{\mathscr{L}\in \mathcal{L}} (-1)^{\kappa(\mathscr{L})} \prod_{(x\rightarrow y) \in \mathscr{L}} (\tilde{\mathbf{q}} - \xi\mathbbm{1})_{yx},
\label{eq:coeffthm}
\end{equation}
where   $\sum_{\mathscr{L}\in \mathcal{L}}$ is a sum over the set $\mathcal{L}$ of all   spanning linear subgraphs $\mathscr{L}$ of the graph represented by $(\tilde{\mathbf{q}}(a) - \xi \mathbbm{1})$,  and $\prod_{(x\rightarrow y) \in \mathscr{L}}$ is a product over all directed edges $(x \rightarrow y)$ that are part of $\mathscr{L}$ (notice that by our choice of convention, the weight associated with the edge going from node $x$ to note $y$ is the entry in the $y$-th row and $x$-th column of the matrix $(\tilde{\mathbf{q}}(a) - \xi \mathbbm{1})$), and where $\kappa(\mathscr{L})$ is the number of  connected components in $\mathscr{L}$.

A {\it linear subgraph} is  a  subgraph for which the indegree  and outdegree of all nodes equals  $1$. Thus, $\mathscr{L}$ is the disjoint union of linear directed cycles, which we differentiate into the following three types:
\begin{enumerate}
    \item cycles consisting of one edge, i.e., $(x\rightarrow x) \in \mathbb{S}_{\mathscr{L}}$ (self loops);
    \item cycles consisting of two directed edges, i.e., $\{(x\rightarrow y),(y \rightarrow x)\} \in \mathbb{E}_{\mathscr{L}}$;
    \item cycles that have three or more directed edges, which we denote by $\mathscr{C} \in  \mathbb{C}_{\mathscr{L}}$, where $\mathscr{C} $ is the set of the directed edges contained in the cycle.
\end{enumerate}
By disjoint we mean that the above mentioned components do not have vertices in common, and we can write this informally as $\mathscr{L} = \mathbb{C}_{\mathscr{L}} \cup \mathbb{S}_{\mathscr{L}}\cup \mathbb{E}_{\mathscr{L}}$ with $\mathbb{C}_{\mathscr{L}}$, $\mathbb{S}_{\mathscr{L}}$ and $\mathbb{E}_{\mathscr{L}}$ mutually disjoint sets.    A linear subgraph  $\mathscr{L}$ is {\it spanning} if all nodes of the original graph are contained in $\mathscr{L}$. 

Having partitioned  spanning linear subgraphs $\mathscr{L}$ into three types of components, we can rewrite the sum in  \eqref{eq:coeffthm} as 
\begin{eqnarray}
   \lefteqn{ {\rm det}\left( \tilde{\mathbf{q}}(a) - \xi \mathbbm{1} \right)}&& \nonumber\\ 
   && = \sum_{\mathscr{L}}  (-1)^{\vert\mathcal{X}\vert+\kappa(\mathscr{L})} 
  \left(\prod_{{(x\rightarrow x)\in\mathbb{S}_\mathscr{L}}}\left(\tilde{\mathbf{q}}_{xx} - \xi\right)\right) \left(\prod_{{\left\{(x\rightarrow y),(y\rightarrow x)\right\}\in\mathbb{E}_\mathscr{L}}}\tilde{\mathbf{q}}_{yx}\tilde{\mathbf{q}}_{xy} \right) \left(\prod_{\mathscr{C}\in\mathbb{C}_{\mathscr{L}}}\prod_{(x\rightarrow y)\in \mathscr{C}}\tilde{\mathbf{q}}_{yx}\right). \nonumber\\
    \label{eq:sumpartitioned}
\end{eqnarray}

\subsection{Coefficients of directed cycles}\label{subsec:S5BB}

The directed cycles $\mathscr{C}$ that have more than three edges are defined by an ordered sequence $\mathscr{C} = [x_1,x_2,\ldots, x_{n(\mathscr{C})},x_1]$ of nodes in $\mathcal{X}$, such that $\tilde{\mathbf{q}}_{x_{i}x_{i+1}} \neq 0$ for all $i$ and each vertex is only traversed once.  Notice that directed cycles are similar to cycles in undirected graphs, as defined  in Sec~\ref{sebsec:fundamentalbasis}, but  reversing the order of the nodes in a cycle $\mathscr{C}$ does not give the same cycle, as  we are considering a directed graph.

We associate to each directed cycle $\mathscr{C}$ the coefficient  
\begin{equation}
c_{\mathscr{C}} = \sum_{(x \rightarrow y) \in \mathscr{C}} c_{xy}. \label{eq:cCDef}
\end{equation}
In what follows, we show that the coefficients $c_{\mathscr{C}}$ can be expressed as a linear combination of the coefficients $c_{\gamma}$ with $\gamma\in \mathcal{C}$, the fundamental cycle basis.    To this aim, we associate to each $\gamma$ two directed cycles, $\mathscr{C}^+_\gamma$ and $\mathscr{C}^-_\gamma$, which traverse the nodes of $\gamma$ in opposing directions. 
It then follows from the definition (\ref{eq:cCDef}) and the fact that $c_{xy}=-c_{yx}$ that 
\begin{equation}
    c_{\mathscr{C}_\gamma^+} = -c_{\mathscr{C}_\gamma^-} = c_\gamma, \label{eq:cgamma1}
\end{equation}
where we use the convention that $c_{\mathscr{C}_\gamma^+} = c_\gamma$.        In fact,  Eq.~(\ref{eq:cgamma1}) extends to any cycle of a graph, and since $\mathcal{C}$ is a fundamental cycle basis, any cycle is a symmetric difference of the cycles $\gamma\in \mathcal{C}$.  Therefore, it holds 
 \begin{equation}
     c_{\mathscr{C}} = \sum_{\gamma \in \mathcal{C}} \epsilon_{\mathscr{C},\gamma} c_\gamma,
     \label{eq:coeffDecomp}
 \end{equation}
 where $\epsilon_{\mathscr{C},\gamma} \in \{-1,0,1\}$ are used to  direction in which the cycles $\gamma$ are traversed.  

\subsection{Characteristic polynomial of $\tilde{\mathbf{q}}(a)$ as a function of the $c_\gamma$}\label{subsec:S5B} 

We  use the definition of $\tilde{\mathbf{q}}(a)$, given by  Eq.~\eqref{eq:qtilde} in the main text,  in the expression (\ref{eq:sumpartitioned}) for the characteristic polynomial. 
We consider each of the three products of Eq.~(\ref{eq:sumpartitioned}) separately, and then we  put them together.

For the self loops we get
\begin{equation}
  \prod_{{(x\rightarrow x)\in\mathbb{S}_\mathscr{L}}}\left(\tilde{\mathbf{q}}_{xx} - \xi\right)=\prod_{{(x\rightarrow x)\in\mathbb{S}_\mathscr{L}}}\left(\mathbf{q}_{xx} - \xi\right),
    \label{eq:sterms}
\end{equation}
which does not depend on the $c_{xy}$.

For the cycles of length two, we get 
\begin{eqnarray}
\prod_{{\left\{(x\rightarrow y),(y\rightarrow x)\right\}\in\mathbb{E}_\mathscr{L}}}\tilde{\mathbf{q}}_{yx} \tilde{\mathbf{q}}_{xy} &=& \prod_{{\left\{(x\rightarrow y),(y\rightarrow x)\right\}\in\mathbb{E}_\mathscr{L}}} \mathbf{q}_{xy} \exp{(-ac_{xy})}\mathbf{q}_{yx} \exp{(-ac_{yx})} 
\nonumber\\ 
&=& \prod_{{\left\{(x\rightarrow y),(y\rightarrow x)\right\}\in\mathbb{E}_\mathscr{L}}}\mathbf{{q}}_{yx}\mathbf{q}_{xy},
    \label{eq:eterms}
\end{eqnarray}
where we have used $c_{xy}  = -c_{yx}$, and hence also this product  does not depend on $c_{xy}$.

Lastly, we consider the directed cycles of length three or larger.    In this case we get
\begin{equation}
   \prod_{\mathscr{C}\in\mathbb{C}_{\mathscr{L}}}\prod_{(x\rightarrow y)\in \mathscr{C}}\tilde{\mathbf{q}}_{yx} =   \prod_{\mathscr{C}\in\mathbb{C}_{\mathscr{L}}}\left( \exp\left(-a c_{\mathscr{C}}\right)\prod_{(x\rightarrow y)\in \mathscr{C}}\mathbf{q}_{yx}\right),
\label{eq:cterms}
\end{equation}
where we have used the definition (\ref{eq:cCDef}) for $c_{\mathscr{C}}$.  

Putting the Eqns.~\eqref{eq:eterms},\eqref{eq:sterms}, and \eqref{eq:cterms} together, we can write the characteristic polynomial as 
\begin{eqnarray}
{\rm det}\left( \tilde{\mathbf{q}}(a) - \xi \mathbbm{1} \right)
&=& \sum_{\mathscr{L}} (-1)^{\vert\mathcal{X}\vert+\kappa(\mathscr{L})}   \left(\prod_{{(x\rightarrow x)\in\mathbb{S}_\mathscr{L}}}\left(\mathbf{q}_{xx} - \xi\right)\right) \left(\prod_{{\left\{(x\rightarrow y),(y\rightarrow x)\right\}\in\mathbb{E}_\mathscr{L}}}\mathbf{{q}}_{yx}\mathbf{{q}}_{xy} \right) 
\nonumber \\&& \qquad \qquad  \times  \left(
\prod_{\mathscr{C}\in\mathbb{C}_{\mathscr{L}}} \exp\left(\mathcal{A}_\mathscr{C}\right)\exp\left(-ac_\mathscr{C}\right) \right), \label{eq:intermediate}
\end{eqnarray}
where we have defined the  ``semi" affinity  $\mathcal{A}_\mathscr{C} = \ln \prod_{(x\rightarrow y)\in\mathscr{C}} \mathbf{q}_{yx} $.
Lastly, using \eqref{eq:coeffDecomp} in \eqref{eq:intermediate}    yields 
\begin{equation}
\begin{split}
{\rm det}\left( \tilde{\mathbf{q}}(a) - \xi \mathbbm{1} \right)= \sum_{\mathscr{L}}  (-1)^{\vert\mathcal{X}\vert+\kappa(\mathscr{L})} \left(\prod_{\mathbb{S}_{(x,x)\in\mathscr{L}}}\left(\mathbf{q}_{xx} - \xi\right)\right)\left(\prod_{{\left\{(x\rightarrow y),(y\rightarrow x)\right\}\in\mathbb{E}_\mathscr{L}}}\mathbf{{q}}_{yx}\mathbf{{q}}_{xy} \right)  \\\left(\prod_{\mathscr{C}\in\mathbb{C}_{\mathscr{L}}} \exp\left(\mathcal{A}_\mathscr{C}\right)\exp\left(-a\sum_{\gamma \in \mathcal{C}} \epsilon_{\mathscr{C},\gamma} c_\gamma\right) \right).
\end{split} \label{eq:detlast}
\end{equation} 
From (\ref{eq:detlast}) we conclude that the characteristic polynomial is a function of the coefficients $c_{\gamma}$ with $\gamma\in \mathcal{C}$.

Notice that if  $J_t=S_t$, the stochastic entropy production, then  $c_\mathscr{C} = \mathcal{A}_\mathscr{C} - \mathcal{A}_\mathscr{\Tilde{C}}$, where $\mathscr{\Tilde{C}}$ is reverse of cycle $\mathscr{C}$, and thus $c_\mathscr{C} = - c_\mathscr{\Tilde{C}}$.    Using these coefficients \eqref{eq:intermediate} readily yields the Galavotti-Cohen symmetry~\cite{lebowitz1999gallavotti},  see also Ref.~\cite{baratoSymmetryCurrentProbability2012} for a detailed analysis.

\section{Model parameters for Kinesin-1 model}\label{sec:fig3params}
The model of Figure~\ref{fig:4}  is the  mechanochemical model for kinesin-1 from Ref.~\cite{shen2022mechanochemical} with one minor modification, namely, we eliminated a unidirectional transition to an absorbing state, corresponding to the detachment of the motor from the biofilament.   The graph in Fig.~\ref{fig:4}(a)  shows the nonzero offdiagonal entries of the $\mathbf{q}$-matrix.  Functional dependence of the rates as a function of the concentration  ${\rm[ATP]}$ of adenosine triphosphate (ATP), the mechanical force $f$ opposing motion of the motor towards the positive end of the microtubule, the Botlzmann constant $k_{\rm B}$, the temperature $\mathsf{T}_{\rm env}$, and the so-called load distribution factors $\delta$ are given by 
\begin{itemize}[itemsep=-2pt]
    \item $\mathbf{q}_{12}=\mathbf{q}^0_{12} {\rm [ATP]}$,
    \item$\mathbf{q}_{13}=\mathbf{q}^0_{13} {\rm exp}\left( \delta_{13}\frac{fd}{k_{\rm_B}\mathsf{T}_{\rm env}}\right)$,
    \item$\mathbf{q}_{15}=\mathbf{q}^0_{15}$,
    \item$\mathbf{q}_{21}=\mathbf{q}^0_{21}$,
    \item$\mathbf{q}_{23}=\mathbf{q}^0_{23} \exp{\left(-\delta_{23}\frac{fd}{k_B \mathsf{T}_{\rm env}}\right)}$,
    \item$\mathbf{q}_{24}=\mathbf{q}^0_{24} \exp{\left(\delta_{24}\frac{fd}{k_{\rm_B}\mathsf{T}_{\rm env}}\right)}$,
    \item$\mathbf{q}_{25}=\mathbf{q}^0_{25}$,
    \item$\mathbf{q}_{31}=\mathbf{q}^0_{24} \exp{\left(-\delta_{31}\frac{fd}{k_{\rm_B}\mathsf{T}_{\rm env}}\right)}$,
    \item$\mathbf{q}_{32}=\mathbf{q}^0_{32} \exp{\left(\delta_{32}\frac{fd}{k_{\rm_B}\mathsf{T}_{\rm env}}\right)}$,
    \item$\mathbf{q}_{34}=\mathbf{q}^0_{34} {\rm [ATP]}$,
    \item$\mathbf{q}_{42}=\mathbf{q}^0_{42} \exp{\left(-\delta_{42}\frac{fd}{k_{\rm_B}\mathsf{T}_{\rm env}}\right)}$,
    \item$\mathbf{q}_{43}=\mathbf{q}^0_{43}$,
    \item$\mathbf{q}_{51}=\mathbf{q}^0_{51}$,
    \item$\mathbf{q}_{52}=\mathbf{q}^0_{52}$.
\end{itemize}

The numerical values chosen for the above parameters are 
\begin{itemize}[itemsep=-2pt]
 \item$\mathbf{q}^0_{34}=\mathbf{q}^0_{12}=9.827 \, \mu M^{-1} s^{-1}$  ,
 \item$\mathbf{q}^0_{43}=\mathbf{q}^0_{21}=5047.875 \, s^{-1}$,
 \item$\mathbf{q}^0_{23}=1627.099\, s^{-1}$,
 \item$\mathbf{q}^0_{32}=0.006\, s^{-1}$,
 \item$\mathbf{q}^0_{24}=\mathbf{q}^0_13=1.666\, s^{-1}$,
 \item$\mathbf{q}^0_{42}=\mathbf{q}^0_{31}=137.582\, s^{-1}$,
 \item$\mathbf{q}^0_{15}=4.344\, s^{-1}$,
 \item$\mathbf{q}^0_{51}=345.215 \, s^{-1}$,
 \item$\mathbf{q}^0_{25}=5146.371 \, s^{-1}$,
 \item$\mathbf{q}^0_{52}=77.252 \, s^{-1}$,
 \item${\rm \left[ATP\right]}=1 \mu M$.
\end{itemize}
The temperature $\mathsf{T}_{\rm env}=296 \,K$  and the Boltzmann constant $k_{\rm_B} = 1.3806\times 10^{-23} \,J/K$.

The load distribution parameters $\delta_{xy}$  are given by $\delta_{13}=0.529$,
$\delta_{31}=0$,
$\delta_{23}=0.055$,
$\delta_{32}=0.416$,
$\delta_{42}=0.006$,
$\delta_{24}=0.523$.  Note that these parameters sum  to one in a cycle, i.e, $\delta_{24} + \delta_{42} + \delta_{23} + \delta_{32} = \delta_{13} + \delta_{31} + \delta_{23} + \delta_{32} = 1$.   

The positional current $J_t$ of the kinesin-1 motor protein is a fluctuating current  determined by the coefficients 
\begin{itemize}
    \item $c_{13} = -c_{31} = (\delta_{13} + \delta_{31})$,  
    \item $c_{24} = -c_{42} = (\delta_{24} + \delta_{42})$, 
    \item and $c_{23} = -c_{32} = (\delta_{23} + \delta_{32})$,
\end{itemize} 
with all other coefficients $c_{xy}$ set to zero.  
One unit of distance travelled by the motor protein  equals the length of one microtubule dimer, which is  $d = 8 \,nm$.      Hence, the distance travelled by kinesin-1 measured in nanometers is given by $J_td$.    
In Figure~\ref{fig:4}(b), the non-dimensionalised version of the mechanical force $\frac{fd}{\rm k_b T_{env}}$ varies along the x-axis. Here, $d$, ${\rm k_B}$ and $\rm T_{env}$ are kept constant at the values mentioned above, while the mechanical force $f$ is varied. With these parameters, the effective affinity $a^\ast$ and $\overline{j}$ were calculated by numerically diagonalizing $\tilde{\mathbf{q}}(a)$ and $\mathbf{q}$ respectively.

\section{Difference between the effective affinity and the stalling force for currents that are not  thermodynamically  consistent}\label{sec:nothermocons}
Figure~\ref{fig:4} of Section~\ref{sec:stallForce}
shows that  thermodynamically consistent currents that satisfy the conditions~\eqref{eq:thermocons} have an  effective affinity that is  approximately equal to the stalling force.    In this Appendix, we show that for currents that are not thermodynamically consistent, the difference between the effective affinity and the stalling force can be large.  

\begin{figure}[t!]
 \centering
 \includegraphics[width=0.7\textwidth]{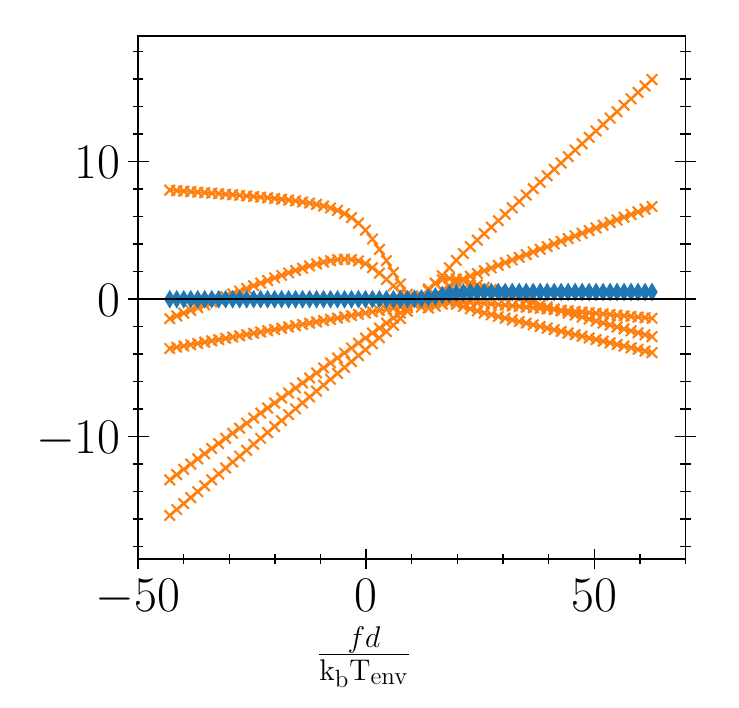}
 \caption{The yellow crosses in the plot above show ($a^\ast - \mathfrak{f}_{\rm stall}$) (i.e, the difference between the left and right hand sides of Eq.~\eqref{eq:toverify}) as a function of $f$, when thermodynamic consistency is broken in the model of Kinesin-1 depicted in Fig.~\ref{fig:4}(a). The different curves are for values of $c_{31}$, $c_{42}$ and $c_{25}$ chosen randomly from a uniform distribution between $0$ and $1$. All other parameters are the same as listed in Appendix~\ref{sec:fig3params}. The blue diamonds show the same plot for the thermodynamically consistent case (as depicted in Fig.~\ref{fig:5}) for comparison.}
\label{fig:6}   
\end{figure}

Indeed, in Fig.~\ref{fig:6} we plot the difference between effective affinity $a^\ast$ and the stalling force $\mathfrak{f}_{\rm stall}$ as as function of the force $f$ for randomly chosen currents that violate the thermodynamic consistency condition in the model of Kinesin-1. As the figure shows, in contrast with the thermodynamically consistent case,   the effective affinity differs significantly  from the stalling force    for  currents that are not thermodynamically consistent.

\end{appendices}

\clearpage


\end{document}